\def\aa{A\&A }					
\def\aas{A\&AS }				
\def\apj{ApJ }					
\def\apjs{ApJS }				
\def\mnras{MNRAS }				
\def\araa{ARA\&A }				
\def\rvmp{RvMP }				
\def\aar{Astron. Astrophys. rev. }
\def\lsim{\lower.5ex\hbox{$\; \buildrel < \over \sim \;$}}
\def\gsim{\lower.5ex\hbox{$\; \buildrel > \over \sim \;$}}
\def\fdg{\hbox{$.\!\!^\circ$}}
\documentclass{aa}
\usepackage{amssymb}
\usepackage{amsmath}
\usepackage{multirow}
\usepackage {graphicx}

\begin{document}


\title{Spatial distribution of interstellar gas
in the innermost 3~kpc of our Galaxy}

\author{K.~Ferri\`ere\inst{1}
		 \and W.~Gillard\inst{2}
		 \and P.~Jean\inst{2}}

\offprints{K.~Ferri\`ere}

\institute{
$^{1}$ LATT-OMP, CNRS/UPS, 31028 Toulouse Cedex 4, France \\
$^{2}$ CESR-OMP, CNRS/UPS, B.P.~4346, 31028 Toulouse Cedex 4, France \\
}

\date{Received  ; accepted }

\titlerunning{Spatial distribution of interstellar gas
in the Galactic bulge}
\authorrunning{K.~Ferri\`ere et al.}


\abstract
{}
{We review the present observational knowledge on the spatial distribution 
and the physical state of the different (molecular, atomic and ionized)
components of the interstellar gas in the innermost 3~kpc of our Galaxy
-- a region which we refer to as the interstellar Galactic bulge,
to distinguish it from its stellar counterpart.}
{We try to interpret the observations in the framework of recent dynamical 
models of interstellar gas flows in the gravitational potential of
a barred galaxy.}
{Relying on both the relevant observations and their theoretical
interpretation, we propose a model for the space-averaged density
of each component of the interstellar gas in the interstellar 
Galactic bulge.}
{}


\maketitle


\section{\label{s1}Introduction}

The state of knowledge on the Galactic center (GC) environment prevailing
one decade ago was reviewed independently by Morris \& Serabyn (1996)
and by Mezger et al. (1996).
To summarize the main points relevant to our work,
the Galactic bulge (GB) is the region of our Galaxy inside
Galactocentric radius $r \simeq 3$~kpc -- a radius which roughly
corresponds to the inner boundary of the Galactic disk molecular ring.
The outer parts of the GB, outside $r \sim 1.5$~kpc,
are largely devoid of interstellar gas.
Atomic gas is confined to a noticeably tilted layer extending 
(in projection) out to $r \sim 1.5$~kpc 
(Burton \& Liszt 1978; Liszt \& Burton 1980).
This H{\sc i} layer is sometimes referred to as the H{\sc i} nuclear disk 
(Morris \& Serabyn 1996), but we find this denomination misleading, 
as we think that the term ``nuclear" should be reserved to objects and
regions with smaller Galactic radii (e.g., $r \lesssim 300$~pc, 
the approximate radius of the nuclear bulge).
For want of any better term, we will refer to this H{\sc i} layer as
the H{\sc i} GB disk.
The GB disk also includes significant amounts of H$_2$,
whose general distribution and kinematics were argued to be similar to
those of H{\sc i} (Liszt \& Burton 1978; Burton \& Liszt 1992;
but see also Sanders et al. 1984; Combes 1991 for a different viewpoint).
However, molecular gas tends to concentrate in the so-called 
central molecular zone (CMZ), an asymmetric layer of predominantly
molecular gas extending (in projection) out to $r \sim 200$~pc --
more exactly, $r \sim 250$~pc at positive longitudes and 
$r \sim 150$~pc at negative longitudes.
The CMZ itself contains a ring-like feature with mean radius $\sim 180$~pc,
now known as the 180-pc molecular ring,
and, deeper inside, a thin sheet populated by dense molecular clouds, 
known as the GC disk population or the GC molecular disk
(Bally et al. 1987, 1988).
The observed kinematics of the 180-pc molecular ring indicate 
strongly non-circular motions, which in turn suggest that either the ring 
is radially expanding (Kaifu et al. 1972; Scoville 1972;
hence the historically often used denomination of ``expanding molecular ring") 
or, more likely, that the gas travels along highly elongated orbits 
(e.g., Bally et al. 1988; Binney et al. 1991).

A number of significant advances were made in the last decade, 
which contributed to improving our knowledge and understanding 
of the interstellar GB.
Nonetheless, several aspects remain ambiguous or controversial,
and important pieces of the puzzle are still missing, 
mainly due to the severe interstellar extinction along the line of sight
to the GC, to the existence of strongly non-circular motions
and to the lack of accurate distance information.
Thus, to date, no complete and fully consistent picture has emerged 
from the vast body of existing observational data.

In this paper, we put together a model for the spatial distribution
of interstellar gas in the Galactic region $r \lesssim 3$~kpc, 
based on the observational results summed up by Morris \& Serabyn (1996)
and Mezger et al. (1996),
on more recent observational studies of dust thermal emission,
CO line emission, H{\sc i} 21-cm line emission and absorption
and pulsar dispersion measures, and on recent theoretical investigations 
of interstellar gas dynamics near the GC.
Our model complements that developed by Ferri\`ere (1998, 2001)
for the Galactic disk outside $r \simeq 3$~kpc. 
In section~\ref{s2}, we review the observations pertaining to 
the spatial distribution and the physical state 
of interstellar gas in its molecular, atomic and ionized forms.
In section~\ref{s3}, we discuss a few dynamical models of interstellar gas
subject to the barred gravitational potential of the Galaxy,
and we make the link between the theoretical predictions of these models
and the observational facts reviewed in section~\ref{s2}.
In section~\ref{s4}, we reconcile as well as possible the disparate
observational and theoretical results presented in the preceding sections,
and we incorporate them into a new, consistent model for the space-averaged 
density of interstellar gas in the Galactic region $r \lesssim 3$~kpc.
In section~\ref{s5}, we conclude our study. 

\begin{figure}
\centering
\includegraphics{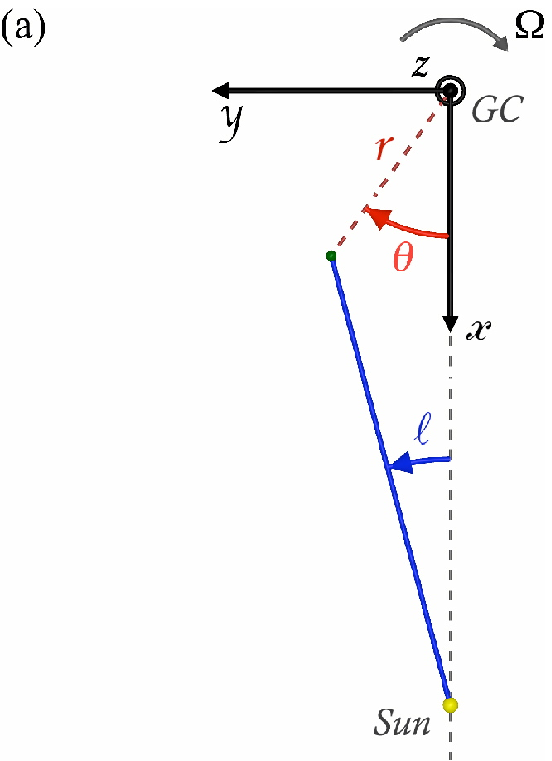} 
\bigskip\bigskip\\
\includegraphics{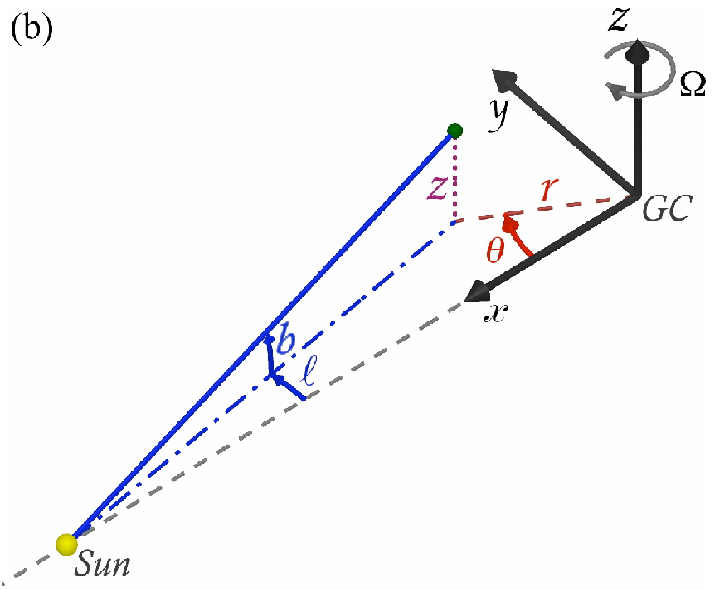}
\caption{\label{fig:coordinates}
The $(x,y,z)$ and $(r,\theta,z)$ Galactocentric coordinate systems 
used in the present paper: (a) two-dimensional view looking down from 
the NGP; (b) full three-dimensional view.
Note that both systems must be left-handed if one wants to stick 
to the usual conventions that $z$ has the same sign as $b$ 
(i.e., positive/negative in the northern/southern Galactic hemisphere) 
and that $\theta$ increases in the direction of Galactic rotation
(i.e., clockwise about the $z$-axis).
}
\end{figure}

For consistency with Ferri\`ere (1998, 2001) and in line with most
papers discussed below, the Sun is assumed to lie at a distance
$r_\odot = 8.5$~kpc from the GC,\footnote{
All our estimates can easily be rescaled to any other value of $r_\odot$ :
distances scale as $r_\odot$, surface densities as $r_\odot^0$, 
volume densities as $r_\odot^{-1}$ and masses as $r_\odot^2$.} 
even though recent work favors a somewhat smaller value.
All spatial distributions will be given either 
as functions of Galactocentric cartesian coordinates $(x,y,z)$,
with the $x$-axis pointing toward the Sun,
the $y$-axis in the direction $l = +90^\circ$
and the $z$-axis toward the North Galactic Pole (NGP),
or as functions of Galactocentric cylindrical coordinates $(r,\theta,z)$,
with $\theta$ increasing in the direction of Galactic rotation, i.e.,
clockwise as seen from the NGP (see Figure~\ref{fig:coordinates}).
Toward the vicinity of the GC, 
$y \simeq (150~{\rm pc}) \ (l / 1^\circ)$ and
$z \simeq (150~{\rm pc}) \ (b / 1^\circ)$, where as usual $l$ and $b$
denote Galactic longitude and latitude, respectively.
Finally, the projected (onto the plane of the sky) horizontal distance 
from the GC is given by
$r_\perp = |y| \simeq (150~{\rm pc}) \ (|l| / 1^\circ)$.

\section{\label{s2}Observational overview}

In this section, we successively review the current observational status
of the total interstellar matter (section~\ref{H}),
the molecular gas (section~\ref{H2}), the atomic gas (section~\ref{HI}) 
and the ionized gas (section~\ref{HII}).
For each, we provide quantitative estimates for 
the space-averaged density and the mass of interstellar hydrogen 
in subregions of the GB.
The density and mass of the total interstellar matter 
(including helium and metals) immediately follow,
once the interstellar elemental abundances in the GB are known.

Various lines of evidence, primarily from H{\sc ii} regions
(Shaver et al. 1983; Afflerbach et al. 1997), planetary nebulae
(Maciel \& Quireza 1999) and early B-type stars (Rolleston et al. 2000),
converge to point to the existence of an inward metallicity gradient
$\sim (0.06 - 0.07)~{\rm dex~kpc}^{-1}$ in the inner Galactic disk.
If extrapolated in to the GC, this metallicity gradient implies 
an increase in metallicity by a factor $\sim 3 - 4$ between the Sun
and the GC.
On the other hand, direct abundance measurements near the GC
yield mixed results : while some confirm the high metallicity
predicted for the GC region (e.g., Smartt et al. 2001), 
others indicate instead a close-to-solar metallicity
(e.g., Carr et al. 2000; Najarro et al. 2004).
Here, we adopt the intermediate, and rather conservative, value 
$Z_{\rm GC} = 2 \, Z_\odot$ for the GC metallicity
(as in Launhardt et al. 2002, and consistent with Sodroski et al. 1995; 
see below).
Furthermore, since the measured He$^+$/H$^+$ ratio does not exhibit 
any significant radial gradient (Shaver et al. 1983), 
we adopt $Y_{\rm GC} = Y_\odot$ for the GC helium fraction.
Altogether, with a helium-to-hydrogen mass ratio of 0.4
and a metal-to-hydrogen mass ratio of $2 \times 0.0265$
(see, e.g., Anders \& Grevesse 1989; D\"appen 2000),
the conversion factor from interstellar hydrogen masses
to total interstellar masses is 1.453.

\subsection{\label{H}Total interstellar matter in the innermost 500~pc}

The large-scale spatial distribution of interstellar matter in the CMZ
was derived by Launhardt et al. (2002), based on IRAS and COBE/DIRBE 
far-infrared maps of the central kpc of the Galaxy.
They converted the optically thin $240~\mu$m dust emission map 
into a hydrogen column density map, using for the dust temperature 
an empirical color temperature proportional to the 60-to-140~$\mu$m 
flux ratio and for the H-to-dust mass ratio the standard value 
near the Sun, $(M_{\rm H} / M_{\rm d})_\odot = 110$, 
divided by the relative metallicity, $(Z_{\rm GC} / Z_\odot) = 2$. 

In contrast to previous authors, Launhardt et al. (2002) defined
the CMZ as the entire region interior to $r \sim 500$~pc,
and they divided it into the inner CMZ, physically associated with
the stellar nuclear bulge ($r \lesssim 230$~pc) and heated by its stars,
and the outer CMZ, containing dust too cold to be possibly heated by 
stars from the nuclear bulge.

\begin{table*}[!htb]
\centering
\caption{\label{tab:MHtot}
Estimated values of the mass of interstellar hydrogen (in any form)
in specific regions of the GB.
        }
\begin{tabular}{llll}
\hline
\hline
Region of space & Radial range & Hydrogen mass & Reference \\
\hline
Inner CMZ : inner disk & 
  $\simeq 0 - 120$~pc & 
  $\sim 4 \times 10^6~M_\odot$ &
  Launhardt et al. (2002) \\
\phantom{Inner CMZ :} outer torus & 
  $\simeq 130~{\rm pc} - 230$~pc &
  $\sim 1.6 \times 10^7~M_\odot$ &
  \qquad " \\
Outer CMZ : $l > 0^\circ$ & 
  $\sim 230~{\rm pc} - 500$~pc &
  $\sim 2.9 \times 10^7~M_\odot$ &
  \qquad " \\
\phantom{Outer CMZ :} $l < 0^\circ$ & &
  $\sim 1.1 \times 10^7~M_\odot$ &
  \qquad " \\
Entire CMZ &
  $\sim 0 - 500$~pc &
  $\sim 6 \times 10^7~M_\odot$ &
  \qquad " \\
\hline
\end{tabular}
\end{table*}

Extinction arguments as well as comparisons with high-resolution submm
continuum maps led Launhardt et al. (2002) to conclude that interstellar
matter in the nuclear bulge is extremely clumpy.
According to them, $\gtrsim 90\%$ of the interstellar matter would be
trapped in small, compact molecular clouds occupying only a few \%
of the interstellar volume. The average hydrogen density inside
these clouds would, therefore, be $\sim$ a few $10^3~{\rm cm}^{-3}$,
with possibly much higher values in the cloud cores.
The remaining $\lesssim 10\%$ of the interstellar matter would form
a diffuse, homogeneously distributed intercloud medium,
Launhardt et al. (2002) further divided the inner CMZ
into a warm inner disk with radius $\simeq 120$~pc
and a cold outer torus centered on the 180-pc molecular ring
and extending radially between $\simeq 130$~pc and 230~pc.
The whole disk/torus structure has a FWHM thickness $\simeq 45$~pc
and a total hydrogen mass $\sim 2.0 \times 10^7~M_\odot$,
with $\sim 4 \times 10^6~M_\odot$ in the inner disk
and $\sim 1.6 \times 10^7~M_\odot$ in the outer torus.
The space-averaged density of hydrogen nuclei can be approximated
by\footnote{The space-averaged densities given by Launhardt et al. (2002)
must be divided by a factor of $3.5$ in order to match the masses and
column densities quoted in their paper (Launhardt, private communication).}
\begin{eqnarray}
\label{eq:Launhardt02_disk}
\langle n_{\rm H} \rangle _{\rm disk} (r,z)
& = &
(72~{\rm cm}^{-3}) \ 
\exp \left[ - \left( \frac{r - 70~{\rm pc}}{L} \right)^4
     \right] \
\nonumber \\
& & 
\times \
\exp \left[ - \left( \frac{|z|}{H} \right)^{1.4} 
     \right]
\end{eqnarray}
in the inner disk and
\begin{eqnarray}
\label{eq:Launhardt02_torus}
\langle n_{\rm H} \rangle _{\rm torus} (r,z)
& = &
(106~{\rm cm}^{-3}) \ 
\exp \left[ - \left( \frac{r - 180~{\rm pc}}{L} \right)^4
     \right] \
\nonumber \\
& & 
\times \
\exp \left[ - \left( \frac{|z|}{H} \right)^{1.4} 
     \right]
\end{eqnarray}
in the outer torus,
with $L = (100~{\rm pc}) / (2 \, (\ln 2)^{1/4}) \simeq 55~{\rm pc}$
and $H = (45~{\rm pc}) / (2 \, (\ln 2)^{1/1.4}) \simeq 29~{\rm pc}$.
Note that strictly speaking, Equation~\ref{eq:Launhardt02_disk}
implies that the inner disk extends radially between $\simeq 20$~pc
and 120~pc, whereas the actual matter distribution is likely to continue
in to the origin; but since the region $r \lesssim 20$~pc contributes very
little to the total mass, this approximation is perfectly valid.

Launhardt et al. (2002) also found large amounts of interstellar dust 
too cold to belong to the nuclear bulge, although its projected
distribution onto the plane of the sky would place most of this dust
inside the nuclear bulge.
The associated interstellar hydrogen has a mass 
$\sim 2.9 \times 10^7~M_\odot$ at positive longitudes
($0\fdg5 \le l \le 4^\circ$)
and $\sim 1.1 \times 10^7~M_\odot$ at negative longitudes
($-4^\circ \le l \le -2^\circ$).
It could be this interstellar matter, located in the outer CMZ,
that would be responsible for the observed asymmetry of the CMZ.
Altogether, the total mass of interstellar hydrogen enclosed within
the central kpc of the Galaxy is $\sim 6 \times 10^7~M_\odot$
(see Table~\ref{tab:MHtot}).

with an average hydrogen density $\sim 10~{\rm cm}^{-3}$,
and the strong UV radiation field from the numerous high-mass stars
embedded in the nuclear bulge would cause this thin intercloud medium 
to be both warm and ionized.
As will become apparent in the course of this section, the two above
density estimates are compatible with more direct density estimates 
for molecular gas (see section~\ref{H2}) and for ionized gas 
(see section~\ref{HII}).

\subsection{\label{H2}Molecular gas}

Radio emission lines of interstellar molecules (excluding H$_2$, which
possesses no permitted lines at radio frequencies)
constitue one of the best tools 
to probe interstellar molecular gas in remote Galactic regions,
and chief amongst them is the $^{12}{\rm CO}~(J = 1 \to 0)$ line at 2.6~mm.
Unfortunately, the $^{12}{\rm CO}~(J = 1 \to 0)$ line as a tracer 
of molecular gas is plagued by the considerable uncertainty in
the $^{12}$CO-to-H$_2$ conversion factor, $X_{\rm CO}$, which relates 
the H$_2$ column density, $N_{{\rm H}_2}$, 
to the velocity-integrated intensity of the $^{12}$CO line, 
$W_{^{12}{\rm CO}}$, 
through $N_{{\rm H}_2} = X_{\rm CO} \ W_{^{12}{\rm CO}}$.

Sodroski et al. (1995) estimated the value of $X_{\rm CO}$ 
near the GC by combining the Goddard-Columbia surveys of
$^{12}{\rm CO}~(J = 1 \to 0)$ emission with COBE/DIRBE observations 
at $140~\mu$m and $240~\mu$m. To deduce the H$_2$ column density 
from the far-infrared data, they assumed that the gas-to-dust mass
ratio is inversely proportional to metallicity, $Z$,
and that $Z$ is $\sim 1.5 - 3$ times higher near the GC
than in the inner Galactic disk
($2~{\rm kpc} \lesssim r \lesssim 7~{\rm kpc}$).
Accordingly, they found that $X_{\rm CO}$ is $\sim 3 - 10$ times lower 
near the GC than in the inner Galactic disk, which, together with
$X_{\rm CO} \sim 2.2 \times 10^{20}~{\rm cm^{-2}~K^{-1}~km^{-1}~s}$
in the inner Galactic disk, leads to
$X_{\rm CO} \sim (2-7) \times 10^{19}~{\rm cm^{-2}~K^{-1}~km^{-1}~s}$
near the GC.
The corresponding H$_2$ mass inside the longitude range
$-1\fdg5 \le l \le 3\fdg5$ (projected half-size $\simeq 375$~pc)
is then $\sim (2-6) \times 10^7~M_\odot$.

Arimoto et al. (1996) studied the radial dependence of the conversion 
factor in more detail, by comparing the CO luminosities to the virial
masses of a large number of giant molecular clouds situated at various
distances from the GC. In this manner, they obtained the relation
(rescaled to $r_\odot = 8.5$~kpc)
\begin{equation}
\label{eq:Arimoto96}
X_{\rm CO} (r) \sim
\left( 9 \times 10^{19}~{\rm cm^{-2}~K^{-1}~km^{-1}~s} \right) \
\exp \left( \frac{r}{7.1~{\rm kpc}} \right) \ \cdot
\end{equation}
Note in passing that Equation~\ref{eq:Arimoto96} implies
$X_{\rm CO} (r_\odot) \sim
3 \times 10^{20}~{\rm cm^{-2}~K^{-1}~km^{-1}~s}$.

\begin{table*}[!htb] 
\centering
\caption{\label{tab:XCO}Estimated values of the $^{12}$CO-to-H$_2$ 
conversion factor, $X_{\rm CO} = N_{{\rm H}_2} / W_{^{12}{\rm CO}}$
(expressed in ${\rm cm^{-2}~K^{-1}~km^{-1}~s}$),
both near the GC ($r \sim 0$) and either in the inner Galactic disk 
$\left( r \simeq 2-7~{\rm kpc} \right)$ or in the vicinity of the Sun 
($r \sim r_\odot$).
        }
\begin{tabular}{llll}
\hline
\hline
$X_{\rm CO} (r \sim 0)$ & 
$X_{\rm CO} \left( r \simeq 2-7~{\rm kpc} \right)$ &
$X_{\rm CO} (r \sim r_\odot)$ & Reference \\
\hline
$\sim (2-7) \times 10^{19}$ & $\sim 2.2 \times 10^{20}$ & & 
Sodroski et al. (1995) \\
$\sim 9 \times 10^{19}$ & & $\sim 3 \times 10^{20}$ &
Arimoto et al. (1996) \\
$\sim (2.4-7.2) \times 10^{19}$ & $\sim 3 \times 10^{20}$ & &
Oka et al. (1998) \\
$\sim (2-4) \times 10^{19}$ & & $\sim 1.5 \times 10^{20}$ &
Strong et al. (2004) \\
\hline
\end{tabular}
\end{table*}

The use of the standard virial mass as a tracer of the actual mass
of molecular clouds was called into question by Oka et al. (1998),
who argued that molecular clouds near the GC are not gravitationally
bound, but instead confined by the high external pressure from hot gas
and magnetic fields.
Using the correct expression for the virial mass of pressure-confined
clouds, they recalibrated the CO-luminosity--virial-mass relation 
for GC molecular clouds, whereupon they obtained 
$X_{\rm CO} \sim 2.4 \times 10^{19}~{\rm cm^{-2}~K^{-1}~km^{-1}~s}$
near the GC.
This value of $X_{\rm CO}$ is one order of magnitude lower than the value
$X_{\rm CO} \sim 3 \times 10^{20}~{\rm cm^{-2}~K^{-1}~km^{-1}~s}$
applying to gravitationally-bound molecular clouds in the Galactic disk.
With their new value of $X_{\rm CO}$, Oka et al. (1998) estimated
the H$_2$ mass inside $r \simeq 375$~pc at $\sim 2 \times 10^7~M_\odot$.
However, they pointed out that their derived values were lower limits,
which could underestimate the true conversion factor and H$_2$ mass 
near the GC by up to a factor of 3.

More recently, Strong et al. (2004) showed that an outward radial
gradient in $X_{\rm CO}$ was required to reconcile the predictions of
their cosmic-ray propagation code GALPROP with the $\gamma$-ray profiles
measured by EGRET/COMPTON, when they adopted the molecular and atomic gas
distributions inferred from CO and H{\sc i} surveys and assumed 
that the distribution of cosmic-ray sources follows the observed
distribution of pulsars.
They did not attempt to derive the precise $r$-dependence of $X_{\rm CO}$
implied by the measured $\gamma$-ray profiles, but they showed that a good fit 
to the EGRET data could be obtained with $X_{\rm CO}$ dropping from 
$\sim 1.5 \times 10^{20}~{\rm cm^{-2}~K^{-1}~km^{-1}~s}$ at $r = r_\odot$
to $\sim 4 \times 10^{19}~{\rm cm^{-2}~K^{-1}~km^{-1}~s}$ at $r = 2$~kpc.
Extrapolating their curve $X_{\rm CO} (r)$ in to the origin yields
$X_{\rm CO} \sim (2-4) \times 10^{19}~{\rm cm^{-2}~K^{-1}~km^{-1}~s}$
at $r \to 0$ (see their Figure~2).

All the above estimates of the $^{12}$CO-to-H$_2$ conversion factor 
are summarized in Table~\ref{tab:XCO}.
It is clear that the value of $X_{\rm CO}$ near the GC remains 
fairly uncertain.
More importantly, the very notion that $N_{{\rm H}_2}$ is proportional
to $W_{^{12}{\rm CO}}$ may be questionable (e.g., Dahmen et al. 1998).
Nevertheless, the fact that three completely different methods
to determine $X_{\rm CO}$ near the GC (far-infrared emission from dust, 
virial masses of molecular clouds and $\gamma$-ray emission from cosmic rays)
broadly converge to a common range of values lends some credence 
to the presumed proportionality between $N_{{\rm H}_2}$ and 
$W_{^{12}{\rm CO}}$.
For the following, we adopt $X_{\rm CO} (r \sim 0) 
= 5 \times 10^{19}~{\rm cm^{-2}~K^{-1}~km^{-1}~s}$
as a reference value, which we consider uncertain by a factor $\sim 2$.

Once the value of $X_{\rm CO}$ is known, the H$_2$ spatial distribution 
can in principle be inferred from CO emission measurements.
The overall disposition of CO emission in the Galaxy was brought to light 
by the early large-scale Galactic CO surveys,
which revealed a sharp peak inside a few 100~pc from the GC, 
a deep depression between $\sim 1.5$~kpc and 3~kpc 
and a ring of enhanced emission (now designated the Galactic disk 
molecular ring) between $\sim 3.5$~kpc and 6~kpc
(see Combes 1991 for a review).

Sanders et al. (1984) presented a full-coverage Galactic
$^{12}{\rm CO}~(J = 1 \to 0)$ survey, on the basis of which 
they constructed an axisymmetric model of the H$_2$ distribution 
in the Galaxy.
According to their model (rescaled to $r_\odot = 8.5$~kpc),
molecular gas in the GB is confined to a disk
with radius $\sim 1.3$~kpc and nearly constant thickness $\sim 70$~pc,
seen edge-on from the Sun and tilted counterclockwise by $\sim 7^\circ$
with respect to the Galactic plane\footnote{Here, and in the rest of 
the paper, the Galactic plane is defined as the plane $b = 0^\circ$,
which is slightly different from the plane of the solar circle.}
(so that its midplane passes from negative latitudes in the first quadrant
to positive latitudes in the fourth quadrant).
This tilted disk is prolonged by a rarefied molecular layer 
with approximately the same thickness, which smoothly joins up with 
the Galactic disk molecular ring.
Taking a constant conversion factor 
$X_{\rm CO} = 3.6\times 10^{20}~{\rm cm^{-2}~K^{-1}~km^{-1}~s}$
for the whole Galaxy, Sanders et al. (1984) found that 
the H$_2$ space-averaged density could be approximated by
\begin{equation}
\label{eq:Sanders84}
\langle n_{\rm H_2} \rangle ({\bf r})
\ = \
\langle n_{\rm H_2} \rangle_{\rm _{0}} (r) \
\exp \left[ - \frac{1}{2} \left( \frac{z - z_0(r,\theta)}{H} \right)^2 
     \right] \ ,
\end{equation}
where (after rescaling to $r_\odot = 8.5$~kpc)
$H \simeq (70~{\rm pc}) / (2 \, \sqrt{2 \, \ln 2}) \simeq 30~{\rm pc}$
out to $r \simeq 3$~kpc,
$z_0(r,\theta)$ is the local Galactic height of the H$_2$ midplane
(given by the tilted disk geometry inside 1.3~kpc)
and $\langle n_{\rm H_2} \rangle_{\rm _{0}} (r)$ is the H$_2$ space-averaged
density at $z_0$, such that
$\langle n_{\rm H_2} \rangle_{\rm _{0}} (0 \to 450~{\rm pc}) 
\simeq 74~{\rm cm}^{-3}$,
$\langle n_{\rm H_2} \rangle_{\rm _{0}} (450~{\rm pc} \to 1.3~{\rm kpc}) 
\simeq 12~{\rm cm}^{-3}$
and $\langle n_{\rm H_2} \rangle_{\rm _{0}} (1.3 \to 3~{\rm kpc}) 
\simeq 1.1~{\rm cm}^{-3}$.
The corresponding H$_2$ masses inside $r = 450$~pc,
between 450~pc and 1.3~kpc, and in the depression zone between 1.3~kpc
and 3~kpc are listed in Table~\ref{tab:MH2}.

\begin{table*}[!htb]
\centering
\caption{\label{tab:MH2}
Estimated values of the mass of interstellar H$_2$ in specific regions
of the GB.
        }
\begin{tabular}{llll}
\hline
\hline
Region of space & Radial range & H$_2$ mass & Reference \\
\hline
Tilted GB disk : inner part &
  $0 - 450$~pc &
  $\sim 1.75 \times 10^8~M_\odot$ &
  Sanders et al. (1984) \\
\phantom{Tilted GB disk :} outer part &
  $450~{\rm pc} - 1.3$~kpc &
  $\sim 2.1 \times 10^8~M_\odot$ &
  \qquad " \\
Depression zone &
  $1.3 - 3$~kpc &
  $\sim 9 \times 10^7~M_\odot$ &
  \qquad " \\
GC molecular disk & 
  $\simeq 0 - 150$~pc & 
  $\sim 2.4 \times 10^7~M_\odot$ &
  Sofue (1995a) \\
``Expanding molecular ring" &
  straddling $r \simeq 180$~pc &
  $\sim 5 \times 10^6~M_\odot$ &
  Sofue (1995b) \\
GB region $-1\fdg5 \le l \le 3\fdg5$ &
  $\simeq 0 - 375$~pc (off-centered) &
  $\sim (2-6) \times 10^7~M_\odot$ &
  Sodroski et al. (1995) \\
GB region $|l| \le 2\fdg5$ &
  $\simeq 0 - 375$~pc &
  $\sim (2-6) \times 10^7~M_\odot$ &
  Oka et al. (1998) \\
Nuclear bulge &
  $\simeq 0 - 280$~pc (off-centered) &
  Dense gas : $\sim (1.2 - 6.4) \times 10^7~M_\odot$ &
  Dahmen et al. (1998) \\
& &
  Thin gas : $\sim (0.7 - 1.4) \times 10^7~M_\odot$ &
  \qquad " \\
\hline
\end{tabular}
\end{table*}

Burton \& Liszt (1992) presented higher-resolution 
$^{12}{\rm CO}~(J = 1 \to 0)$ observations toward the GC
and pointed out that the central parts of the gaseous GB disk 
are in fact closely aligned with the Galactic plane.
More specifically, they showed that the observed $(l,b,v)$ pattern
of $^{12}$CO emission in the GB could be understood 
in the framework of a tilted and warped H$_2$ disk model with 
the following characteristics (rescaled to $r_\odot = 8.5$~kpc):
Out to $r \simeq 170$~pc, the H$_2$ disk is flat, parallel to 
the Galactic plane and $\simeq 30$~pc thick
(Gaussian scale height $\simeq 13$~pc).
Between $\simeq 170$~pc and 1.5~kpc, the H$_2$ disk flares linearly 
to a thickness $\simeq 300$~pc (uncertain value),
and this flaring is accompanied by a warp such that the H$_2$ midplane
becomes tilted with respect to the Galactic plane, by an angle that varies 
sinusoidally with Galactic azimuth between $-13^\circ$ at $\theta = 45^\circ$ 
(so that $b < 0^\circ$ in the near, first-quadrant sector)
and $+13^\circ$ at $\theta = 225^\circ$
($b > 0^\circ$ in the far, fourth-quadrant sector) (see their Figure 5).
Overall, the warped H$_2$ disk appears tilted out of the Galactic plane 
and inclined to the line of sight, very much like the tilted H{\sc i}
disk of Burton \& Liszt (1978) (see section~\ref{HI}).
The warped H$_2$ disk also bears a resemblance to Sanders et al.'s (1984)
tilted H$_2$ disk, but it differs by its substantial flaring 
and by the direction of maximum tilt ($\theta = 45^\circ$ - $225^\circ$, 
as opposed to $\theta = 90^\circ$ - $270^\circ$ in Sanders et al.).
Interestingly, the H$_2$ space-averaged density in the warped disk 
can be expressed by Equation~\ref{eq:Sanders84}, with 
$H \simeq 13~{\rm pc}$ inside $r \simeq 170$~pc and
$H \simeq 0.088 \ (r - 22~{\rm pc})$ between $\simeq 170$~pc and 1.5~kpc,
and with $\langle n_{\rm H_2} \rangle_{\rm 0} \propto 1/H$
(such that the H$_2$ column density through the disk has a constant, 
albeit unspecified, value).
Besides its geometric parameters, the warped H$_2$ disk model also
possesses two kinematic parameters describing rotation and expansion 
motions, respectively.

Let us now focus on the innermost Galactic regions, inside a few 100~pc 
from the GC, where molecular gas tends to concentrate.
Sofue (1995a) investigated the H$_2$ morphology and kinematics inside 
$|l| \simeq 1^\circ$ ($r_\perp \simeq 150$~pc), 
based on the AT\&T Bell Laboratories survey of
$^{13}{\rm CO}~(J = 1 \to 0)$ emission conducted by Bally et al. (1987).
He adopted
$X_{\rm CO} = 9.2 \times 10^{19}~{\rm cm^{-2}~K^{-1}~km^{-1}~s}$
(from a slightly modified version of Equation~\ref{eq:Arimoto96}) 
together with $W_{^{12}{\rm CO}} / W_{^{13}{\rm CO}} = 6.2$
(measured intensity ratio averaged over $|l| \lesssim 20^\circ$).
The most salient continuous features in the longitude-velocity $(l,v)$
diagrams of the GC molecular disk are two apparently rigidly-rotating
ridges. Sofue (1995a) interpreted these ridges as the $(l,v)$ traces of
two dense material arms,
which together would form a rotating ring with mean radius $\simeq 120$~pc 
and vertical thickness $\sim 15$~pc (except for the massive molecular 
complexes around Sgr~B and C, whose thickness reaches $\sim 30 - 60$~pc).
An analysis of the velocity-integrated intensity maps yields
a total H$_2$ mass $\sim 2.4 \times 10^7~M_\odot$ for the disk,
with $\sim 1.1 \times 10^7~M_\odot$ in Arm~I
and $\sim 0.8 \times 10^7~M_\odot$ in Arm~II,
and an H$_2$ mass $\sim 2.8 \times 10^7~M_\odot$ for the entire region 
$r_\perp \lesssim 150$~pc.
By examining a somewhat wider section of Bally et al.'s (1987) survey,
Sofue (1995b) estimated that the ``expanding molecular ring"
that surrounds the GC molecular disk
has a mean radius $\simeq 180$~pc (as already established before), 
a radial thickness $\sim 15$~pc, a vertical thickness $\sim 100$~pc 
and an H$_2$ mass $\sim 5 \times 10^6~M_\odot$
(80\% of which belongs to the region $r_\perp \lesssim 150$~pc).

Dahmen et al. (1998) gained additional information on the spatial
structure and physical state of molecular gas near the GC by analyzing 
a SMWT C$^{18}{\rm O}~(J = 1 \to 0)$ survey 
of the central region $-1\fdg05 \le l \le 3\fdg6$,
in conjunction with the SMWT $^{12}{\rm CO}~(J = 1 \to 0)$ survey
of the broader region $-12^\circ \le l \le 13^\circ$
presented by Bitran et al. (1997).
For both isotopomers,\footnote{Isomers having the same number of
isotopically distinct atoms but differing in the positions of these atoms. 
The term "isotopomer" results from the contraction of the words "isotope" 
and "isomer".}
they performed radiative transfer calculations
in the Large Velocity Gradient (LVG) approximation, and
they adopted abundance ratios $^{12}{\rm CO} / {\rm H}_2 = 10^{-4}$
and C$^{18}{\rm O} / {\rm H}_2 = 4 \times 10^{-7}$.
Their calculations enabled them to estimate the kinetic temperature,
volume density and total mass of molecular gas in the nuclear bulge,
which they assumed to extend over the longitude range
$-1\fdg5 \le l \le 2\fdg25$ (projected half-size $\simeq 280$~pc).
They found that most of the C$^{18}$O emission is likely to originate 
in molecular gas with kinetic temperature $T \sim 50$~K and
H$_2$ density $n_{{\rm H}_2} \sim 10^{3.5}~{\rm cm}^{-3}$,
and they argued that the associated H$_2$ mass in the nuclear bulge
is comprised between $\sim 1.2 \times 10^7~M_\odot$ and
$6.4 \times 10^7~M_\odot$.
Furthermore, they noted significant differences between 
the C$^{18}$O and $^{12}$CO intensity maps, which they interpreted
as evidence for the presence of a widespread, high-temperature and
low-density gas component, detected in $^{12}$CO but not in C$^{18}$O.
This thin gas would have $T \sim 150$~K,
$n_{{\rm H}_2} \sim 10^{2.5}~{\rm cm}^{-3}$
and an H$_2$ mass in the nuclear bulge between 
$\sim 0.7 \times 10^7~M_\odot$ and twice that value.

Another powerful method to probe the physical conditions in molecular gas
near the GC is to compare emission lines from different $J$ levels of
$^{12}$CO. In this spirit, Oka et al. (1998) mapped the region
$|l| \le 2\fdg5$ 
in $^{12}{\rm CO}~(J = 2 \to 1)$ emission with the NRO survey telescope,
and they resorted to previous $^{12}{\rm CO}~(J = 1 \to 0)$ maps
from Bitran (1987) for comparison.
Their analysis of the $^{12}{\rm CO}~(J = 2 \to 1)$ /
$^{12}{\rm CO}~(J = 1 \to 0)$ intensity ratio indicates that the CO
luminosity of the innermost $\sim 400$~pc is dominated by emission
from low-density gas with $n_{{\rm H}_2} \sim 10^{2.5}~{\rm cm}^{-3}$.
To reconcile this finding with the inference from CS, NH$_3$ and HCN
observations that high-density gas with 
$n_{{\rm H}_2} \gtrsim 10^{4}~{\rm cm}^{-3}$ prevails,
Oka et al. (1998) advanced the view that molecular gas actually exists
in two distinct components: 
a low-density ($n_{{\rm H}_2} \lesssim 10^{3}~{\rm cm}^{-3}$)
component present in ``diffuse" clouds with a large filling factor,
and a high-density ($n_{{\rm H}_2} \gtrsim 10^{4}~{\rm cm}^{-3}$)
component present in ``clumps" with a small filling factor.
This view of a two-component molecular gas in the GC region has found
support in several subsequent studies relying on other tracer molecules 
(e.g., Rodr\'iguez-Fern\'andez et al. 2001; Oka et al. 2005; 
Magnani et al. 2006).
Let us simply mention that the warm ($T \sim 150$~K), and presumably
low-density, component detected by Rodr\'iguez-Fern\'andez et al. (2001)
would on average represent $\sim 30\%$ of the total molecular gas.

\begin{table*}[!htb]
\centering
\caption{\label{tab:MH2bis}
Estimated values of the mass of interstellar H$_2$ in specific regions
of the GB,
deduced from $^{12}$CO measurements (or from $^{13}$CO measurements 
combined with an assumed $^{12}$CO-to-$^{13}$CO intensity ratio)
together with $X_{\rm CO} = 5 \times 10^{19}~{\rm cm^{-2}~K^{-1}~km^{-1}~s}$.
        }
\begin{tabular}{llll}
\hline
\hline
Region of space & Radial range & H$_2$ mass & Original reference$^a$ \\
\hline
Tilted GB disk : inner part &
  $0 - 450$~pc &
  $\sim 2.4 \times 10^7~M_\odot$ &
  Sanders et al. (1984) \\
\phantom{Tilted GB disk :} outer part &
  $450~{\rm pc} - 1.3$~kpc &
  $\sim 2.9 \times 10^7~M_\odot$ &
  \qquad " \\
Depression zone &
  $1.3 - 3$~kpc &
  $\sim 1.3 \times 10^7~M_\odot$ &
  \qquad " \\
GC molecular disk & 
  $\simeq 0 - 150$~pc & 
  $\sim 1.3 \times 10^7~M_\odot$ &
  Sofue (1995a) \\
``Expanding molecular ring" &
  straddling $r \simeq 180$~pc &
  $\sim 2.7 \times 10^6~M_\odot$ &
  Sofue (1995b) \\
GB region $-1\fdg5 \le l \le 3\fdg5$ &
  $\simeq 0 - 375$~pc (off-centered) &
  $\sim 4.4 \times 10^7~M_\odot$ &
  Sodroski et al. (1995) \\
GB region $|l| \le 2\fdg5$ &
  $\simeq 0 - 375$~pc &
  $\sim 4.2 \times 10^7~M_\odot$ &
  Oka et al. (1998) \\
\hline
\multicolumn{4}{l}{\tiny $^a$ Reference for the corresponding mass estimate 
given in Table~\ref{tab:MH2}, which we rescaled to our adopted value of 
$X_{\rm CO}$.}
\end{tabular}
\end{table*}

Higher-$J$ transitions of $^{12}$CO also provide a valuable diagnostic
tool for molecular gas. For instance, Martin et al. (2004)
surveyed the region $-1\fdg3 \le l \le 2^\circ$ 
in $^{12}{\rm CO}~(J = 4 \to 3)$ and $^{12}{\rm CO}~(J = 7 \to 6)$ 
emission with the AST/RO telescope, and they analyzed their
$^{12}{\rm CO}~(J = 4 \to 3)$ and $^{12}{\rm CO}~(J = 7 \to 6)$ maps
together with existing $^{12}{\rm CO}~(J = 1 \to 0)$ and
$^{13}{\rm CO}~(J = 1 \to 0)$ maps from Stark et al. (1988) and
Bally et al. (1987, 1988).
Like Dahmen et al. (1998), they worked with the LVG approximation
and they took $^{12}{\rm CO} / {\rm H}_2 = 10^{-4}$.  
What emerges from their study is that the molecular gas kinetic temperature 
decreases from $T \gtrsim 70$~K at the edges of cloud complexes
to $T \lesssim 50$~K in their interiors,
and that the H$_2$ density spans the validity range of their LVG analysis 
($\sim 10^{2.5} - 10^{4.5}~{\rm cm}^{-3}$), with typical values 
$n_{{\rm H}_2} \sim 10^{3.5}~{\rm cm}^{-3}$ inside cloud complexes.

Finally, Sawada et al. (2004) devised a purely observational method
to deconvolve sky maps of CO emission
into a face-on view of the molecular gas distribution near the GC,
based on a quantitative comparison between 
the 2.6-mm $^{12}$CO emission line and the 18-cm OH absorption line.
Their face-on map exhibits a strong central condensation
(corresponding to the CMZ), which is elongated along an axis
inclined by $\sim 70^\circ$ to the line of sight
(so that its near end lies at positive longitudes),
$\sim 500~{\rm pc} \times 200~{\rm pc}$ in size, 
and possibly composed of a pair of arms (as suggested by Sofue 1995a).
The so-called ``expanding molecular ring" lies at the periphery 
of the central condensation, without clearly separating from it.
It, too, is elongated and inclined (by an uncertain angle $< 70^\circ$)
toward positive longitudes, though it does not actually stand out 
as a coherent entity in the face-on map.
Instead, Sawada et al.'s (2004) results tend to substantiate the idea
that the ``expanding molecular ring" arises from the projection of 
interstellar gas moving along highly elongated orbits 
in response to the gravitational potential of the Galactic bar 
(see section~\ref{s3} for further details).

Table~\ref{tab:MH2} provides a summary of 
all the H$_2$ masses estimated in this subsection.
Clearly, the H$_2$ mass derived by Sanders et al. (1984) 
for the innermost region is significantly greater than the other estimates. 
This is only because these authors used a much larger value 
of the conversion factor, $X_{\rm CO}$, more typical of the Galactic disk 
at large.
If we now rescale all the H$_2$ masses deduced from $^{12}$CO
measurements (or from $^{13}$CO measurements combined with 
an assumed $^{12}$CO-to-$^{13}$CO intensity ratio) in Table~\ref{tab:MH2} 
to our adopted reference value 
$X_{\rm CO} = 5 \times 10^{19}~{\rm cm^{-2}~K^{-1}~km^{-1}~s}$,
we obtain the H$_2$ masses listed in Table~\ref{tab:MH2bis}.
These masses agree with each other to within a factor of $\sim 2$.
Moreover, they grossly match the total hydrogen masses inferred 
from infrared dust measurements by Launhardt et al. (2002) 
(see Table~\ref{tab:MHtot}), consistent with the notion that most of 
the interstellar gas in the CMZ is in molecular form.

\subsection{\label{HI}Atomic gas}

Virtually all our observational knowledge on the interstellar atomic gas 
near the GC stems from spectral studies of the H{\sc i} 21-cm line 
in emission or in absorption. 
The early NRAO surveys of H{\sc i} 21-cm emission over the longitude range 
$-11^\circ \le l \le 13^\circ$ led Burton \& Liszt (1978) and 
Liszt \& Burton (1980) to propose a tilted disk model for the H{\sc i} 
spatial distribution inside $r \sim 2$~kpc.
The model parameters were determined iteratively by generating synthetic 
21-cm emission spectra and adjusting them to the observed spectra.

In Burton \& Liszt's (1978) model (rescaled to $r_\odot = 8.5$~kpc),
the H{\sc i} disk is axisymmetric, 1.3~kpc in radius, 200~pc thick
(Gaussian scale height of 85~pc), tilted by $22^\circ$ out of the Galactic 
plane (with the first/fourth quadrant side at negative/positive latitudes) 
and inclined by $78^\circ$ to the plane of the sky (with the near/far side 
at negative/positive latitudes),
such that the normal to the H{\sc i} disk forms a $25^\circ$ angle
with the $z$-axis.
The H{\sc i} space-averaged density in the disk is given by
a Gaussian function of distance to the H{\sc i} midplane, $z_d$,
\begin{equation}
\label{eq:Burton78}
\langle n_{\rm H{\scriptscriptstyle I}} \rangle (z_d)
\ = \
(0.39~{\rm cm}^{-3}) \
\exp \left[ - \frac{1}{2} \left( \frac{z_d}{85~{\rm pc}} \right)^2 
     \right] \ ,
\end{equation}
which implies a total H{\sc i} mass in the disk
$\simeq 1.1 \times 10^7~M_\odot$.
Finally, the H{\sc i} gas within the disk has rotation and expansion
motions of comparable magnitudes.

Liszt \& Burton (1980) argued that the observed kinematics of 
the H{\sc i} gas could be more plausibly explained by a tilted
elliptical disk model, in which the gas moves along closed
elliptical orbits.
Thus, in Liszt \& Burton's (1980) model (again rescaled
to $r_\odot = 8.5$~kpc), the H{\sc i} disk is elliptical, with semi-major
axis 1.6~kpc and axis ratio $3.1:1$.
Its thickness remains 200~pc, but the tilt and inclination angles are now
$13\fdg5$ to the Galactic plane and $70^\circ$ to the plane of the sky
(entailing a $24^\circ$ angle between the disk normal and the $z$-axis),
and the major axis is oriented at $48\fdg5$ clockwise 
to the ``inclined $x$-axis" (corresponding to a $51\fdg5$ angle 
to the line of sight) (see their Figure 1).
The H{\sc i} space-averaged density in the disk is still given by
Equation~\ref{eq:Burton78}, but the total H{\sc i} mass in the disk
now amounts to $\simeq 5.2 \times 10^6~M_\odot$.

\begin{table*}[!htb]
\centering
\caption{\label{tab:MHI}
Estimated values of the mass of interstellar H{\sc i} in specific
regions of the GB.
        }
\begin{tabular}{llll}
\hline
\hline
Region of space & Radial range & H{\sc i} mass & Reference \\
\hline
Tilted GB disk : & & & 
  \\
\qquad Axisymmetric &
  $0 - 1.3$~kpc &
  $\sim 1.1 \times 10^7~M_\odot$ &
  Burton \& Liszt (1978) \\
\qquad Elliptical &
  $0 - 1.6$~kpc / $0 - 0.52$~kpc&
  $\sim 5.2 \times 10^6~M_\odot$ &
  Liszt \& Burton (1980) \\
\qquad Axisymmetric &
  See Equation~\ref{eq:Burton93} &
  $\sim 2.2 \times 10^6~M_\odot$ &
  Burton \& Liszt (1993) \\
Inner part of GB disk &
  $\sim 0 - 300$~pc (off-centered) & 
  $\sim 2.3 \times 10^6~M_\odot$ &
  Rohlfs \& Braunsfurth (1982) \\
180-pc molecular ring &
  straddling $r \simeq 175$~pc (off-centered) &
  $\sim 8 \times 10^5~M_\odot$ &
  \qquad " \\
\hline
\end{tabular}
\end{table*}

A different modification to Burton \& Liszt's (1978) original model 
can be found in the paper by Burton \& Liszt (1993).
There, the H{\sc i} spatial distribution within the GB disk is modeled 
in a slightly more complicated manner, 
with a H{\sc i} space-averaged density depending on both $r$ and $z_d$:
\begin{eqnarray}
\label{eq:Burton93}
\langle n_{\rm H{\scriptscriptstyle I}} \rangle (r,z_d)
& = &
\left\{ 
  (0.82~{\rm cm}^{-3}) \
  \exp \left[ - \left( \frac{r}{450~{\rm pc}} \right)^2 \right]
\right.
\nonumber \\
& &
\left.
+ (0.10~{\rm cm}^{-3}) \
  \exp \left[ - \left( \frac{r}{800~{\rm pc}} \right)^2 \right] 
\right\} \
\nonumber \\
& &
\times \
\exp \left[ - \frac{1}{2} \left( \frac{z_d}{50~{\rm pc}} \right)^2 
     \right] \ \cdot
\end{eqnarray}
The corresponding FWHM thickness is $\simeq 118$~pc
and the total H{\sc i} mass is $\simeq 2.2 \times 10^6~M_\odot$.
However, the emphasis of the paper is placed on the kinematic aspects
of atomic gas in the GB, and as the authors themselves recognize,
Equation~\ref{eq:Burton93} may have little intrinsic significance.
Therefore, we do not regard the associated mass estimate as particularly
trustworthy.

Rohlfs \& Braunsfurth (1982) presented
a more refined description of the morphology and kinematics of atomic
gas in the central parts of the H{\sc i} GB disk.
They analyzed 21-cm emission as well as absorption spectra obtained with
the Effelsberg telescope at longitudes $|l| \le 1\fdg5$, and they made 
comparisons between both types of spectra in order to locate 
the major H{\sc i} features relative to the GC.
Like Burton \& Liszt (1978), they started from the premise that
the velocity field includes rotation and expansion.
Furthermore, they used an axisymmetric rotation curve inferred from
a Galactic mass model as an independent distance indicator,
and they assumed that the major H{\sc i} features are nearly circular
(though not necessarily concentric with the GC).
Because of these stringent assumptions, largely refuted by more recent work, 
too much reality should not be ascribed to the precise quantitative
estimates provided by Rohlfs \& Braunsfurth (1982).
Nevertheless, a couple of important points deserve mentioning.
First, although the H{\sc i} GB disk as a whole is markedly tilted,
its central parts are approximately parallel to the Galactic plane
(as is the case for the warped H$_2$ disk of Burton \& Liszt 1992; 
see section~\ref{H2}).
Second, the 180-pc molecular ring is visible in 21-cm emission, 
without however being prominent.
For reference, Rohlfs \& Braunsfurth (1982) estimated that 
the inner part of the H{\sc i} GB disk has 
a radius $\sim 300$~pc, a thickness $\sim 90$~pc
and an H{\sc i} mass $\sim 2.3 \times 10^6~M_\odot$,
while the 180-pc molecular ring has a mean radius $\simeq 175$~pc, 
a vertical thickness $\sim 60$~pc
and an H{\sc i} mass $\sim 8 \times 10^5~M_\odot$.

Unfortunately, the atomic gas does not benefit from the wide variety 
of diagnostic spectral lines or from the vast number of high-resolution
surveys as are presently available for the molecular gas.
Further genuine progress in our understanding of interstellar atomic gas
near the GC will probably have to await the release of new data
from the GC extension of the Southern Galactic Plane Survey 
(McClure-Griffiths, private communication).
In the meantime, one should bear in mind that the spatial distribution
and kinematics of the atomic gas might not be as drastically different 
from those of the molecular gas as traditionally believed.
Liszt \& Burton (1996) advanced several cogent arguments pointing to
the congruence of both gases in position and in velocity,
and they attributed their apparent disparities to differences 
in the observational conditions (angular resolution, sky coverage 
and signal-to-noise ratio) of H{\sc i} {\it versus} CO,
in the thin {\it versus} thick optical-depth regime 
and in the presence {\it versus} absence of absorption 
(see also Liszt \& Burton 1978).
In addition, they emphasized that the kinematic features appearing in 
the H{\sc i} and CO spectra (notably the ``expanding molecular ring")
do not correspond to discrete material bodies, but instead result from
kinematic projection effects.

Again, we close up the subsection with a summary table
(Table~\ref{tab:MHI}) including all our mass estimates.
For completeness, we can add that the H{\sc i} mass inside
$r \simeq 300$~pc is $\simeq 5.8 \times 10^5~M_\odot$
in Burton \& Liszt's (1978) model and $\simeq 6.6 \times 10^5~M_\odot$
in their (1993) model, as opposed to $\sim 3.1 \times 10^6~M_\odot$ 
in Rohlfs \& Braunsfurth (1982).
Hence, there is roughly a factor of 5 discrepancy between 
Burton \& Liszt on the one hand and Rohlfs \& Braunsfurth 
on the other hand.
The same factor exists, for the entire H{\sc i} GB disk,
between the old and new models of Burton \& Liszt,
but as we already mentioned earlier, the H{\sc i} mass deduced from
their 1993 model should not be taken too seriously.
If we disregard this mass estimate, we may conclude from a comparison 
between Tables~\ref{tab:MH2bis} and \ref{tab:MHI}
that the H{\sc i} mass of the entire GB disk represents but
$\sim 10\% - 20\%$ of its H$_2$ mass.

\subsection{\label{HII}Ionized gas}

Radio signals from pulsars and other (Galactic and extragalactic)
compact sources provide a unique source of information on the ionized
component of the interstellar gas.
Cordes \& Lazio (2002) assembled all the useful data (dispersion
measures, scattering measures and independent distance estimates)
available at the end of 2001 to construct a new non-axisymmetric model 
of the spatial distribution of interstellar free electrons in the Galaxy.
This ``NE2001 model", which incorporates separate multi-wavelength data
on the spiral structure of the Galaxy, the local interstellar medium
and the GC region, supersedes the earlier models of Cordes et al. (1991) 
and Taylor \& Cordes (1993).
It includes a smooth large-scale component, which consists of 
two axisymmetric disks and five spiral arms, 
a smooth GC component, a contribution from the local interstellar medium,
and individual clumps and voids.

Retaining solely the smooth components present inside 3~kpc, we can write
the free-electron space-averaged density in the GB as
\begin{equation}
\label{eq:Cordes05}
\langle n_{\rm e} \rangle ({\bf r})
\ = \
\langle n_{\rm e} \rangle _{\rm 1} (r,z)
+ \langle n_{\rm e} \rangle _{\rm 2} (r,z)
+ \langle n_{\rm e} \rangle _{\rm 3} (r,z) \ ,
\end{equation}
where
\begin{eqnarray}
\label{eq:Cordes05_thick}
\langle n_{\rm e} \rangle _{\rm 1} (r,z)
& = &
(0.05~{\rm cm}^{-3}) \ 
\left[ \cos \left( \pi \ \frac{r}{2 \, L_1} \right) \
       u ( L_1 - r )
\right] \
\nonumber \\
& & 
\times \
{\rm sech}^2 \left( \frac{z}{H_1} \right) \ ,
\end{eqnarray}
with $L_1 = 17$~kpc and $H_1 = 950$~pc, for the outer thick disk 
($u$ is the unit step function), 
\begin{eqnarray}
\label{eq:Cordes05_thin}
\langle n_{\rm e} \rangle _{\rm 2} (r,z)
& = &
(0.09~{\rm cm}^{-3}) \ 
\exp \left[ - \left( \frac{r - L_2}{L_2/2} \right)^2
     \right] \
\nonumber \\
& & 
\times \
{\rm sech}^2 \left( \frac{z}{H_2} \right) \ ,
\end{eqnarray}
with $L_2 = 3.7$~kpc and $H_2 = 140$~pc, for the inner thin disk\footnote{The 
expression 
of $g_2(r)$ given in Table~2 of Cordes \& Lazio (2002) contains a typo: 
$A_a$ should be $A_2$ in the numerator and $A_2/2$ in the denominator
(Lazio, private communication).}, and
\begin{eqnarray}
\label{eq:Cordes05_GC}
\langle n_{\rm e} \rangle _{\rm 3} (r,z)
& = &
(10~{\rm cm}^{-3}) \ 
\exp \left[ - \frac{x^2 + (y - y_3)^2}{L_3^2} 
     \right]
\nonumber \\
& & 
\times \
\exp \left[ - \frac{(z - z_3)^2}{H_3^2}
     \right] \ ,
\end{eqnarray}
with $y_3 = -10$~pc, $z_3 = -20$~pc, $L_3 = 145$~pc and $H_3 = 26$~pc,
for the GC component.\footnote{The $(x,y,z)$ coordinates used by 
Cordes \& Lazio (2002) correspond to $(y,x,z)$ in our coordinate system
(see Figure~\ref{fig:coordinates}).}
Strictly speaking, the offset ellipsoid exponential in 
Equation~\ref{eq:Cordes05_GC} should be truncated to zero for arguments 
smaller than $-1$. However, such a discontinuity was introduced into the model
to reflect the abrupt changes observed in the distribution of scattering 
diameters for OH masers in the GC rather than an abrupt drop-off
in the free-electron density. Since we are interested only in 
the free-electron density (as opposed to both the density and
its fluctuations), we are entitled to relax the constraint arising from 
scattering diameters and omit to truncate the ellipsoid exponential 
in Equation~\ref{eq:Cordes05_GC} (Lazio, private communication).
Then the GC component remains dominant near the midplane out to 
$r \simeq 335$~pc. The rest of the GB is largely dominated
by the thick disk, except for a narrow wedge starting at 
$(r,z) \simeq (2.3~{\rm kpc},0)$ and opening outward,
wherein the thin disk takes over.

To convert the above free-electron density
into an ionized-hydrogen density, we follow conventional wisdom 
and assume that the molecular and atomic media (discussed in 
sections~\ref{H2} and \ref{HI}, respectively) can be considered 
as completely neutral,
while the ionized medium can be divided into a warm ionized medium (WIM),
where hydrogen is entirely ionized and helium completely neutral, 
and a hot ionized medium (HIM), where hydrogen and helium are both 
fully ionized.
If we further denote by $f_{\rm HIM}$ the fraction of ionized gas
belonging to the HIM and remember that helium represents 10\% by number 
of hydrogen (see beginning of section~\ref{s2}), we find that 
the ionized-hydrogen space-averaged density is related to 
the free-electron space-averaged density through
\begin{equation}
\label{eq:n_e}
\langle n_{\rm H^+} \rangle =
\frac{1}{1 + 0.2 \ f_{\rm HIM}} \
\langle n_{\rm e} \rangle \ \cdot
\end{equation}
The HIM fraction, $f_{\rm HIM}$, is quite uncertain,
but fortunately, its exact value has little impact on the inferred 
H$^+$ density, $\langle n_{\rm H^+} \rangle$, which varies between 
$\langle n_{\rm e} \rangle$ if $f_{\rm HIM} = 0$
and $0.83 \ \langle n_{\rm e} \rangle$ if $f_{\rm HIM} = 1$.

The total H$^+$ mass in the interstellar GB and the contributions 
from the thick disk, thin disk and GC component are obtained by integrating
Equations~\ref{eq:Cordes05_thick}, \ref{eq:Cordes05_thin} and
\ref{eq:Cordes05_GC}, divided by the factor $(1 + 0.2 \ f_{\rm HIM})$,
out to $r = 3$~kpc.
For reference, if $f_{\rm HIM} = 0$,
the thick disk has an H$^+$ mass of 
$6.4 \times 10^7~M_\odot$ inside 3~kpc, 
with $1.8 \times 10^6~M_\odot$ inside 500~pc
and $1.6 \times 10^5~M_\odot$ inside 150~pc;
the thin disk has an H$^+$ mass of 
$8.3 \times 10^6~M_\odot$ inside 3~kpc, 
with $1.8 \times 10^4~M_\odot$ inside 500~pc
and $1.0 \times 10^3~M_\odot$ inside 150~pc;
and the GC component has a total H$^+$ mass of 
$7.5 \times 10^5~M_\odot$,
with $5.0 \times 10^5~M_\odot$ inside 150~pc
and negligible amounts outside 500~pc.
The resulting total H$^+$ masses inside 150~pc, 500~pc and 3~kpc
are displayed in Table~\ref{tab:MHII}.
A comparison with our previous tables (Table~\ref{tab:MHtot} 
or Table~\ref{tab:MH2bis} with an assumed H{\sc i}-to-H$_2$ mass ratio
$\sim 10\% - 20\%$) suggests that ionized gas accounts for only
$\sim 4\% - 5\%$ of the total interstellar gas in the region 
$r \lesssim 150$~pc, $\sim 4\% - 8\%$ in the region $r \lesssim 500$~pc, 
and $\sim 30\% - 50\%$ on average over the entire GB
($r \lesssim 3$~kpc).
Not surprisingly, the fraction of ionized gas increases not only with
$|z|$ (as indicated by the comparatively large scale height in
Equation~\ref{eq:Cordes05_thick}), but also with $r$.

\begin{table*}[!htb]
\centering
\caption{\label{tab:MHII}
Estimated values of the mass of interstellar H$^+$ (total mass or partial 
contribution from the WIM or HIM/VHIM) in specific regions of the GB.
The total H$^+$ masses deduced from the NE2001 model for the free-electron
density (Cordes \& Lazio 2002) are those obtained under the extreme
assumption that 100\% of the ionized medium is warm
(so that $\langle n_{\rm H^+} \rangle = \langle n_{\rm e} \rangle$).
If, in reality, a fraction $f_{\rm HIM}$ of the ionized medium is hot,
these masses should be divided by the factor $(1 + 0.2 \ f_{\rm HIM})$
(see Equation~\ref{eq:n_e}).
        }
\begin{tabular}{llllll}
\hline
\hline
Region of space & Radial range & \multicolumn{3}{c}{H$^+$ mass} & Reference \\
& & \qquad Total & \qquad WIM & \quad HIM/VHIM & \\
\hline
Galactic bulge : innermost &
  $0 - 150$~pc & 
  $\sim 6.6 \times 10^5~M_\odot$ & & &
  Cordes \& Lazio (2002) \\
\phantom{Galactic bulge :} intermediate &
  $0 - 500$~pc & 
  $\sim 2.6 \times 10^6~M_\odot$ & & &
  \qquad " \\
\phantom{Galactic bulge :} total &
  $0 - 3$~kpc & 
  $\sim 7.3 \times 10^7~M_\odot$ & & &
  \qquad " \\
Smaller ellipsoid &
  $\simeq 0 - 47$~pc & 
  & $\sim 1.2 \times 10^5~M_\odot$ & &
  Mezger \& Pauls (1979) \\
Larger ellipsoid &
  $\simeq 0 - 112$~pc & 
  & $\sim 4.7 \times 10^5~M_\odot$ & &
  \qquad " \\
Galactic bulge &
  $0 - 3$~kpc & 
  & & $\sim 8 \times 10^6~M_\odot$ &
  Snowden et al. (1997) \\
& & & & $\sim 1.2 \times 10^7~M_\odot$ &
  Almy et al. (2000) \\
10-keV plasma : innermost &
  $0 - 150$~pc & 
  & & $\sim 5.5 \times 10^4~M_\odot$ &
  \raisebox{-6pt}{$\Big\{ $}\begin{tabular}[t]{l}
                            \!\!\!Yamauchi et al. (1990) \\
                            \!\!\!Koyama et al. (1996)
                            \end{tabular}
  \\
\phantom{10-keV plasma :} total &
  $\sim 0 - 250$~pc & 
  & & $\sim 9.5 \times 10^4~M_\odot$ &
  \qquad " \\
\hline
\end{tabular}
\end{table*}

Lazio and Cordes (1998) discussed the nature of the medium responsible
for the observed scattering. They proposed that scattering arises in
thin layers on the surfaces of molecular clouds.
These layers would either be the photoionized skins of molecular clouds,
with $T_e \sim 10^4$~K and $n_e \gtrsim 10^3~{\rm cm}^{-3}$,
or the interfaces between molecular clouds and the hot ($\sim 10^7$~K) 
ambient gas, 
with $T_e \sim (10^5 - 10^6)$~K and $n_e \sim (5 - 50)~{\rm cm}^{-3}$.
In both cases, the scattering medium would have a small filling factor.
However, the scattering medium possibly constitutes but a small part of 
the ionized medium; therefore, its morphology and physical parameters 
are not necessarily representative of the ionized medium in general.

The best source of information on the morphology and physical parameters
of the warm component of the ionized medium near the GC comes from the thermal 
(free-free) component of the radio continuum emission and from radio 
recombination lines.
In general, thermal continuum measurements are used to image the ionized gas
and to estimate its electron density and temperature,
while recombination line measurements make it possible to investigate 
the kinematics of the ionized gas and also to estimate its electron 
temperature.

As explained in the review by Mezger \& Pauls (1979),
the thermal radio continuum emission is produced in an extended
ionized medium (here referred to as the WIM)
and in individual H{\sc ii} regions.
On the sky, the thermal radio emission appears closely correlated 
with far-infrared dust emission, and less well correlated with CO emission, 
which is asymmetric (see section~\ref{H2}), more extended in longitude 
and less extended in latitude.
To quote Mezger \& Pauls (1979), ``this suggests that the [WIM] may be 
ionization bounded along the Galactic plane, but density bounded 
perpendicular to it."
As a first rough approximation, Mezger \& Pauls (1979) modeled the WIM by
the superposition of two oblate ellipsoids with the following
characteristics\footnote{
In the rest of this section, all density estimates deduced from emission
measures implicitly assume that the phase under consideration occupies 
all the interstellar volume. If the considered phase actually has 
a volume filling factor $\phi$, each of our density estimates corresponds 
in fact to the geometric mean between the true and space-averaged densities,
and the actual true/space-averaged density is obtained 
by dividing/multiplying our density estimate by $\sqrt{\phi}$.}
(rescaled to $r_\odot = 8.5$~kpc):
the larger ellipsoid is $\simeq (225~{\rm pc})^2 \times 90~{\rm pc}$ in size
and has $T_e \simeq 5000$~K, $n_e \simeq 8~{\rm cm}^{-3}$ and
$M_{\rm H^+} \simeq 4.7 \times 10^5~M_\odot$, while 
the smaller ellipsoid is $\simeq (95~{\rm pc})^2 \times 55~{\rm pc}$ in size
and has $T_e \simeq 5000$~K, $n_e \simeq 18~{\rm cm}^{-3}$ and
$M_{\rm H^+} \simeq 1.2 \times 10^5~M_\odot$ (see Table~\ref{tab:MHII}).
Altogether, the central electron density is $\simeq 26~{\rm cm}^{-3}$
and the total H$^+$ mass is $\simeq 5.9 \times 10^5~M_\odot$.
For comparison, in the NE2001 model of Cordes \& Lazio (2002),
the GC component has an ellipsoid exponential distribution
with FWHM size $\simeq (241~{\rm pc})^2 \times 43~{\rm pc}$ 
(comparable in radial extent, but significantly more oblate than 
the larger ellipsoid of Mezger \& Pauls 1979), 
a central electron density of $10~{\rm cm}^{-3}$ and a total H$^+$ mass of 
$(7.5 \times 10^5~M_\odot) / (1 + 0.2 \ f_{\rm HIM})$.
Evidently, the good agreement found for the H$^+$ masses is a little
fortuitous.

A number of studies on the WIM near the GC have been carried out
since the review of Mezger \& Pauls (1979).
For instance, Mehringer et al. (1992, 1993) observed the Sgr B region
with the VLA, both at several wavelengths in the radio continuum 
and in the H110$\alpha$ radio recombination line.
They estimated the r.m.s. electron density outside compact H{\sc ii} regions 
at $\sim 80~{\rm cm}^{-3}$ in Sgr B1
and $\sim 60~{\rm cm}^{-3}$ in Sgr B2.

Besides  the WIM, the ionized medium near the GC contains a hot
component (referred to as the HIM) which can be detected through 
its X-ray thermal emission.
Maps of the diffuse X-ray background in the $(0.5 - 2.0)$~keV energy
band from the ROSAT all-sky survey reveal an extended zone of enhanced
emission in the general direction of the GC (Snowden et al. 1997).
Although part of the enhancement can be attributed to the nearby Loop~I SB,
Snowden et al. (1997) argued, based on the latitude profile of the excess 
emission, on the absorption trough running along the Galactic plane
and on the deep shadows cast by relatively distant molecular clouds,
that the bulk of the enhancement arises from a bulge of hot, X-ray
emitting gas around the GC.
They went on to propose a crude model for the X-ray bulge,
which relies on the $(0.5 - 2.0)$~keV ROSAT data corrected for
foreground absorption by interstellar H{\sc i} and H$_2$,
and which assumes that the X-ray emitting gas is in collisional ionization
equilibrium.
In their model, the hot gas has a temperature of $4 \times 10^6$~K,
an electron space-averaged density (again rescaled to $r_\odot = 8.5$~kpc)
\begin{eqnarray}
\label{eq:Snowden97}
\langle n_{\rm e} \rangle _{\rm _{HIM}} (r,z)
& = &
(0.0034~{\rm cm}^{-3}) \ 
u ( 6~{\rm kpc} - r ) \
\nonumber \\
& & 
\times \
\exp \left[ - \left( \frac{|z|}{2~{\rm kpc}} \right)
     \right] \ ,
\end{eqnarray}
and hence an H$^+$ mass $\simeq 8 \times 10^6~M_\odot$ inside $r = 3$~kpc.

As follow-up work, Almy et al. (2000) analyzed the shadows cast 
by a distant molecular cloud complex against the X-ray enhancement 
in the $\frac{3}{4}$~keV and 1.5~keV ROSAT bands,
and came to the conclusion that a significant fraction of the observed
radiation in that direction is indeed emitted beyond the cloud complex,
most likely in an X-ray bulge surrounding the GC.
They also derived a simple, albeit more realistic, model for the X-ray bulge,
in which the X-ray emitting gas is an adiabatic polytrope 
($P \propto \rho^{5/3}$) in hydrostatic balance (in the Galactic
gravitational potential of Wolfire et al. 1995).
Like Snowden et al. (1997), they assumed collisional ionization equilibrium 
and they accounted for foreground absorption.
They found that the best fit to the $\frac{3}{4}$~keV ROSAT data is reached 
for a central gas temperature of $8.2 \times 10^6$~K
and a central electron density of $0.011~{\rm cm}^{-3}$.

Almy et al. (2000) provided no analytical expressions to describe 
the spatial variation of the hot gas parameters. Nonetheless, all the useful 
expressions can be retrieved by solving the hydrostatic equation,
$-{\bf \nabla} P - \rho \ {\bf \nabla} \phi = 0$,
together with the polytropic assumption, $P = K \, \rho^{5/3}$,
and Wolfire et al.'s (1995) representation of the Galactic
gravitational potential,
\begin{eqnarray}
\label{eq:Wolfire95}
\phi (r,z)
& = & 
- (225~{\rm km~s}^{-1})^2 \
\nonumber \\
& \times &
\left\{
\frac{C_1}
     {\sqrt{r^2 + \left( a_1 + \sqrt{z^2 + b_1^2} \right)^2}}
\ + \ \frac{C_2}{a_2 + \sqrt{r^2 + z^2}}
\right.
\nonumber \\
& &
\left.
\quad - \ C_3 \ 
\ln \frac{\displaystyle \sqrt{1 + \frac{a_3^2 + r^2 + z^2}
                         {r_h^2}
               } - 1
         }
         {\displaystyle \sqrt{1 + \frac{a_3^2 + r^2 + z^2}
                         {r_h^2}
               } + 1
         }
\right\} \ ,
\end{eqnarray}
where $C_1 = 8.887$~kpc, $a_1 = 6.5$~kpc, $b_1 = 0.26$~kpc, 
$C_2 = 3.0$~kpc, $a_2 = 0.70$~kpc, $C_3 = 0.325$, $a_3 = 12$~kpc
and $r_h = 210$~kpc.
In this manner, it is found that the electron space-averaged density
of the HIM can be written in the form
\begin{eqnarray}
\label{eq:Almy00}
\langle n_{\rm e} \rangle _{\rm _{HIM}} (r,z)
& \! = \! &
\Big\{ \!
(0.011~{\rm cm}^{-3})^{2/3}
- (1.73 \times 10^{-17}~{\rm cm}^{-4}~{\rm s}^2)
\nonumber \\
& & 
\quad \times \
\big[ \phi(r,z) - \phi(0,0) \big]
\Big\} ^{1.5}
\end{eqnarray}
and that the H$^+$ mass inside $r = 3$~kpc (with $|z| \le 5$~kpc)
amounts to $\simeq 1.2 \times 10^7~M_\odot$.

Hence, the mass of hot H$^+$ predicted by Almy et al.'s (2000) model
for the region $r \le 3$~kpc is 1.5 times that predicted by 
Snowden et al.'s (1997) cruder model.
In view of Table~\ref{tab:MHII} and the discussion below Equation~\ref{eq:n_e}, 
Almy et al.'s  model suggests that hot gas globally represents 
$\sim 17\%$ of all the ionized gas, and thus $\sim 5\% - 8.5\%$ 
of the total interstellar gas, in the GB.
However, little can be said about the spatial variation of the hot gas
fraction, because the assumption of an adiabatic polytrope
in hydrostatic balance is made for modeling convenience rather than 
on observational grounds.

The ionized medium near the GC also contains an even higher-temperature
component (sometimes called the very hot ionized medium or VHIM), 
revealed by its hard X-ray thermal emission, both in a free-free continuum 
and in characteristic lines from highly ionized elements.
Koyama et al. (1989) reported the detection with the Ginga satellite 
of intense 6.7-keV K$\alpha$ line emission from helium-like iron,
the most abundant ionization state at a temperature of several keV.
From the shape of the (supposedly free-free) continuum in the $(2 - 18)$~keV 
energy band, they independently estimated the plasma temperature at 
$k T \sim 10$~keV (i.e., $T \sim 10^8$~K). 
Yamauchi et al. (1990) then performed a two-dimensional mapping of 
the very hot 6.7-keV emitting plasma (outside the Galactic ridge).
They found that its two-dimensional distribution on the plane of the sky
could be fitted by an elliptical Gaussian with FWHM size 
$\simeq 270~{\rm pc} \times 150~{\rm pc}$ and major axis tilted clockwise 
by $\simeq 21^\circ$ with respect to the Galactic plane.

Subsequent, higher--energy-resolution observations in the $(2 - 10)$~keV
band with the ASCA satellite confirmed the presence of very hot plasma
in the GC region (Koyama et al. 1996).
Although the X-ray continuum could again be explained by thermal
free-free emission at $k T \gtrsim 10$~keV,
the simultaneous appearance in the spectra of K$\alpha$ lines from helium-like 
and hydrogen-like ions of various elements suggests that the very hot plasma 
is either multi-temperature or out of ionization equilibrium.
Its electron density
was estimated by Koyama et al. (1996) at $\sim (0.3 - 0.4)~{\rm cm}^{-3}$.
When integrated over an ellipsoid Gaussian distribution with FWHM size
$\simeq (270~{\rm pc})^2 \times 150~{\rm pc}$ (from Yamauchi et al. 1990),
this electron density implies a total H$^+$ mass 
$\sim (8 - 11) \times 10^4~M_\odot$, 58\% of which lies inside $r = 150$~pc.
From this, we may conclude that the very hot gas encloses $\sim 7\% - 10\%$ 
of the ionized gas, and hence $\sim 0.3\% - 0.5\%$ of all 
the interstellar gas, inside $r = 150$~pc.


\section{\label{s3}Dynamical models}

The observed non-axisymmetric distribution and non-circular motions 
of interstellar gas near the GC can most likely be attributed to 
the presence of a Galactic bar, and to a large extent, they can be 
explained in terms of the theoretical properties of stable closed orbits 
in the gravitational potential of a barred galaxy 
(see Binney \& Merrifield 1998 for a brief review).
Both analytical and numerical calculations of particle orbits 
in a barred gravitational potential have brought to light the existence 
of two families of stable closed orbits.
Although the exact status of these two families depends on the shape and
strength of the bar and can become rather complex, for our purposes 
it suffices to be aware of the basic typical situation. 
Typically, the orbits prevailing inside the bar's inner Lindblad resonance 
(ILR) are the so-called $x_2$ orbits, elongated perpendicular to the bar's 
major axis, and the orbits prevailing outside the ILR are 
the $x_1$ orbits, generally elongated along the bar 
-- except between the bar's corotation radius and its outer Lindblad 
resonance (OLR), where most closed orbits are unstable
(see, e.g., Contopoulos \& Mertzanides 1977;\footnote{
It is interesting to note that the $x_1$ {\it versus} $x_2$ terminology
in use today is opposite to that initially introduced by 
Contopoulos \& Mertzanides (1977) in the case of a weak bar.} 
Contopoulos \& Papayannopoulos 1980; Athanassoula 1992a).
The ellipticity of both families of orbits increases toward the ILR,
and on the whole $x_1$ orbits are more elongated than $x_2$ orbits.

Binney et al. (1991) proposed a dynamical model of interstellar gas 
in the innermost $(2-3)$~kpc of the Galaxy meant to reproduce at best
the observed morphology and kinematics of CO, CS and H{\sc i} emissions.
The basic scenario underlying their model is the following:
Interstellar gas subject to the gravitational potential of the bar
tends to settle onto stable closed orbits, 
although due to diverse dissipation processes, it gradually drifts inward
through a sequence of decreasing-energy orbits.
Gas starting off outside the ILR travels along $x_1$ orbits
until its inward drift brings it to a cusped $x_1$ orbit, interior to which 
$x_1$ orbits become self-intersecting at their apocenters.
Shocks and material collisions along the cusped and outermost
self-intersecting $x_1$ orbits cause the gas to abruptly lose energy and
angular momentum and to drop onto the lower-energy $x_2$ orbits, along which 
it continues its spiraling motion toward the GC.
Since $x_2$ orbits are overall rounder than $x_1$ orbits,
gas loses energy and drifts inward less rapidly in the ``$x_2$ disk",
which therefore attains higher densities.

Binney et al. (1991) interpreted the observational longitude-velocity 
$(l,v)$ diagrams of CO and H{\sc i} emissions in terms of 
the theoretical $(l,v)$ curves generated by non self-intersecting 
$x_1$ orbits and by $x_2$ orbits.
The cornerstone of their discussion is the parallelogram that encloses
most of the emission in the CO $(l,v)$ diagram (this parallelogram 
corresponds to the 180-pc molecular ring discussed earlier, which, 
as we already mentioned, is also known as the ``expanding molecular ring").
Binney et al. argued that the parallelogram could be identified with
the $(l,v)$ trace of the cusped $x_1$ orbit, thereby placing 
the GC molecular disk (i.e., the GC disk population of dense molecular 
clouds) on $x_2$ orbits.
No such parallelogram appears in the H{\sc i} $(l,v)$ diagram,
whose structure suggests instead that atomic gas extends well into 
the region of non self-intersecting $x_1$ orbits.
These considerations led Binney et al. to propose that a substantial
fraction of the interstellar gas turns molecular at the strong shocks 
present in the vicinity of the cusped $x_1$ orbit.
Farther out, the observed gap in the interstellar gas distribution
between radii $\sim 1.5$~kpc and $\sim 3$~kpc could presumably
be explained by the generally unstable character of particle orbits 
in a region extending radially from somewhere inside corotation to the OLR
and by the resulting expulsion of gas away from the unstable region.
The accumulation of expelled gas just outside the OLR would then lie
at the root of the Galactic disk molecular ring.

Detailed quantitative comparisons between the observational $(l,v)$ diagrams
and the theoretical $(l,v)$ curves enabled Binney et al. (1991) 
to place constraints on the parameters of their model.
The best constrained parameter is the inclination angle of the bar's
major axis to the line of sight, which was found to be 
$\theta_{\rm bar} \simeq 16^\circ$ (within a few degrees)
toward positive longitudes.
Much more uncertain are the bar's dimensions; Binney et al. arbitrarily
adopted $a_0 = 1.2$~kpc for the bar's semi-major axis (more exactly, 
the radius along the major axis beyond which the matter's density falls off 
very steeply) and $4:3:3$ for the axis ratios, but the bar could easily be 
longer and more elongated.
The bar's corotation radius was arbitrarily set to $r_{\rm CR} = 2.4$~kpc,
consistent with the notions that the bar terminates somewhat inside 
corotation (Contopoulos \& Papayannopoulos 1980; Sellwood \& Sparke 1988; 
Athanassoula 1992b) and that the region around corotation is depopulated 
of gas.

For all its successes, Binney et al.'s (1991) model leaves a number of
important observational features unexplained.
Most notably, the longitudinal lopsidedness in the observed CO distribution
is much more pronounced than the asymmetry expected from the bar's
inclination toward positive longitudes.
In addition, Binney et al. noted a significant discrepancy in the velocity 
scales of the observational CO and H{\sc i} $(l,v)$ diagrams,
when interpreted in the framework of their model. They attributed
this discrepancy to the neglect of gas-dynamical effects 
in their particle-orbit approach, arguing that hydrodynamical forces 
at the strong shocks responsible for the transition from $x_1$ 
to $x_2$ orbits near the ILR lead to large departures from ballistic
trajectories.

This argument prompted Jenkins \& Binney (1994) to simulate interstellar
gas flows in Binney et al.'s (1991) bar model.
They tested two radically different two-dimensional, sticky-particle schemes, 
which neglect self-gravity, but allow for both accretion of matter shed by
bulge stars and removal of clouds from regions above a given density 
threshold. Both schemes yield similar results, which qualitatively confirm 
the dynamical scenario envisioned by Binney et al., with one important 
difference.
Here it is found that gas parcels approaching (but still outside)
the cusped $x_1$ orbit tumble onto $x_2$ orbits located deeper
inside, so that a nearly empty zone develops near the cusped orbit
between the ``$x_1$ ring" and the ``$x_2$ disk".
Only an external source of matter, such as gas loss by bulge stars,
can keep this transition zone appreciably populated.
For future reference, the ILR lies at $r_{\rm ILR} \simeq  350$~pc,
the cusped orbit has a half-size $\simeq 650~{\rm pc} \times 150~{\rm pc}$,
and for reasonable values of the numerical parameters,
the inner boundary of the $x_1$ ring has a half-size 
$\simeq (750 \pm 50)~{\rm pc} \times (400 \pm 50)~{\rm pc}$
and the $x_2$ disk a half-size
$\simeq (280 \pm 20)~{\rm pc} \times (180 \pm 20)~{\rm pc}$.

Regrettably, Jenkins \& Binney's (1994) simulations fail to resolve 
the main problems with Binney et al.'s (1991) model, namely, 
its inability to reproduce the pronounced lopsidedness in the observed 
CO distribution and its inability to simultaneously explain 
the observational CO and H{\sc i} $(l,v)$ diagrams.
Jenkins \& Binney suggested that part of the blame could probably be
laid on the limitations of their numerical schemes, which are restricted
to two dimensions and omit self-gravity.
They also pointed out that the unsteadiness of gas flows near the ILR
could be partly (though certainly not entirely) responsible for 
the observed asymmetry in CO.
Finally, they raised the possibility that CO and H{\sc i} line emissions
might not be reliable tracers of interstellar gas mass.

Subsequently, many hydrodynamical simulations of gas flows in barred 
gravitational potentials were performed with various numerical techniques.
Englmaier \& Gerhard (1999) employed a two-dimensional smoothed-particle
hydrodynamics code to follow interstellar gas flows in the Galactic
gravitational potential deduced from the deprojected and dust-corrected
COBE/DIRBE near-infrared luminosity distribution (see Binney et al. 1997),
under the assumption of a constant mass-to-light ratio.
The COBE data imply that the bar's inclination angle is restricted to
the interval $15^\circ - 35^\circ$, with a preferred value
$\theta_{\rm bar} = 20^\circ$. With this value, the bar has a semi-major
axis $a \sim 2$~kpc and axis ratios $\simeq 5:3:2$,
it is surrounded by a thin elliptical disk with radius $\sim 3.5$~kpc 
along the major axis and $\sim 2$~kpc along the minor axis,
and corotation occurs at $r_{\rm CR} \simeq  3$~kpc
(assuming $r_\odot = 8$~kpc; Binney et al. 1997).
Englmaier \& Gerhard (1999) supplemented the gravitational potential 
deduced from the COBE near-infrared luminosity with the contributions
from a central cusp and (for some runs) an outer dark halo.
They presented their simulation results in the form of face-on
surface-density maps of interstellar gas in the inner Galaxy ($r < r_\odot$) 
together with the associated $(l,v)$ diagrams.
The maps clearly feature a four-armed spiral pattern outside corotation
and two pairs of arms inside corotation, each pair emanating approximately 
from one end of the bar. Interior to the two innermost arms,
the vicinity of the cusped $x_1$ orbit becomes rapidly depleted, 
as gas parcels reaching the cusped orbit's shock quickly fall in along 
the shock ridges onto the central $x_2$ disk.
The latter stands out in the surface-density maps as a compact disk 
of radius $\sim 150$~pc.

Bissantz et al. (2003) ran similar hydrodynamical simulations 
based on the improved near-infrared luminosity distribution models of 
Bissantz \& Gerhard (2002).
In the best-fit model that they adopted, the bar's inclination angle 
is again $\theta_{\rm bar} = 20^\circ$, but its semi-major axis is now 
$a \simeq 1.75$~kpc, its axis ratios $\simeq 5:2:1.5$ and its corotation
radius $r_{\rm CR} \simeq  3.4$~kpc (assuming again $r_\odot = 8$~kpc).
Bissantz et al. (2003) examined the gas dynamics in the bar region 
more closely.
As a starting point, they computed closed $x_1$ and $x_2$ orbits
in their model gravitational potential and found that the outermost
$x_2$ orbit does not reach out to the cusped $x_1$ orbit.
This, they argued, could be the reason why a gap opens up
in the simulations between the cusped orbit and the $x_2$ disk.
According to their calculations, the cusped orbit has a half-size 
$\simeq 1.35~{\rm kpc} \times 280~{\rm pc}$
(approximately twice as large as in Jenkins \& Binney 1994)
and the outermost closed $x_2$ orbit has a half-size 
$\simeq 175~{\rm pc} \times 85~{\rm pc}$ (nearly twice smaller than 
the half-size of the $x_2$ disk in Jenkins \& Binney 1994).

All the above simulations have two important limitations:
they are two-dimensional and, more critically, the Galactic
gravitational potential is prescribed at the outset and taken to be
time-independent and point-symmetric with respect to the GC.
A major improvement on both counts was brought by Fux (1999),
who performed three-dimensional, self-consistent composite simulations
of the inner Galaxy, in which the stellar bar and the interstellar gas
are evolved in concert, with $N$-body and smoothed-particle
hydrodynamics codes, respectively.
All gravitational interactions (including self-gravity) are taken 
into account, and no stationarity or symmetry condition is imposed.
The simulated face-on surface-density maps are found to fluctuate 
in time. The stellar bar varies in shape and its center of mass 
oscillates around the GC with an amplitude of several 100~pc.
The interstellar gas flow is unsteady and asymmetric; it develops
transient spiral arms and shock fronts as well as a ring or disk
of $x_2$ orbits which closely follows the bar's center of mass.

Fux (1999) constrained the location of the Sun relative to the bar 
with the help of dust-corrected COBE/DIRBE K-band data.
Moreover, he constructed movies of synthetic $(l,v)$ diagrams, 
compared them to the observational CO and H{\sc i} $(l,v)$ diagrams
and selected the best-matching snapshots.
He then used the corresponding gas flow models to interpret the main
features of the observational $(l,v)$ diagrams.
The two preferred models have $\theta_{\rm bar} \simeq 25^\circ$
(compatible with the COBE data) and $r_{\rm CR} \simeq 4.5$~kpc.
In the model that Fux singled out for attention (the one that gives
the best overall agreement with the $(l,v)$ observations), 
the cusped $x_1$ orbit has a half-size 
$\simeq 3.2~{\rm kpc} \times 1.3~{\rm kpc}$, 
which is considerably larger than in the simulations described above. 
With such large dimensions, the cusped orbit cannot possibly correspond to
the parallelogram in the observational CO $(l,v)$ diagram,
as suggested by Binney et al. (1991).
Instead, Fux's conjecture is that the parallelogram could represent
interstellar gas streams originating from the cusped orbit
or from dustlanes (associated with ``off-axis shocks") connecting
the cusped orbit to the $x_2$ ring/disk, and grazing the $x_2$ ring/disk.
Aside from this new interpretation, Fux pointed out that the bar parameters 
cannot be estimated with any confidence from the best-matching models, 
and he devised a model-independent method relying on purely geometric 
considerations.
This method yields $\theta_{\rm bar} = 25^\circ \pm 4^\circ$ for 
the bar's inclination angle, $a \simeq 3.2$~kpc for its semi-major axis,
$a:b \simeq 5:3$ for its face-on axis ratio
and $r_{\rm CR} \simeq  (4.0 \pm 0.5)$~kpc for its corotation radius
(assuming $r_\odot = 8$~kpc).

A great merit of Fux's (1999) study is that it underscores the inherent
non-stationarity of the Galactic bar, with its large-amplitude
oscillations around the GC, and the resulting unsteadiness and 
off-centering of the interstellar gas distribution.
In this manner, it offers a natural explanation for the observed 
longitudinal asymmetry in the CO intensity maps.
Furthermore, since the cusped $x_1$ orbit is no longer associated with
the parallelogram in the CO $(l,v)$ diagram, the difference in 
the velocity scales of the observational CO and H{\sc i} $(l,v)$ diagrams
no longer poses a problem.
On the other hand,  Fux's simulations raise their own set of questions,
as the large values obtained for the corotation radius and 
for the dimensions of the cusped orbit seem difficult to reconcile with 
the observed gas deficit between $r \sim 1.5$~kpc and 3~kpc
and with the Galactic disk molecular ring starting at $r \simeq 3.5$~kpc.

Another, more subtle aspect of the interstellar gas morphology 
near the GC concerns the orientation of the central $x_2$ ring/disk 
with respect to the bar's major axis.
We know that closed particle (i.e., collisionless) orbits in a barred 
gravitational potential 
are elongated either parallel or perpendicular to the bar's major axis.
Typically, they are elongated parallel to the bar between the ILR and
corotation and outside the OLR, and perpendicular to the bar inside 
the ILR (where the central $x_2$ ring/disk resides) and between corotation
and the OLR (e.g., Contopoulos \& Mertzanides 1977; 
Contopoulos \& Papayannopoulos 1980).
This is not true of closed gas (i.e., collisional) orbits. 
Once pressure and viscous forces are taken into account, 
the $90^\circ$ jumps in orbit orientation at the bar's resonances 
are completely smoothed out.
Wada (1994), who studied the case of a weak bar with an analytical
damped-oscillator model, found that the orbit inclination angle 
to the bar's major axis, $\Delta \theta$, varies gradually with radius,
such that $0^\circ < \Delta \theta \le 45^\circ$ between the ILR and corotation 
and outside the OLR, and $45^\circ \le \Delta \theta < 90^\circ$
inside the ILR and between corotation and the OLR.
Hence, $x_2$ orbits lead the bar by $\Delta \theta > 45^\circ$,
and the whole $x_2$ ring/disk should appear inclined to the bar 
by a little more than $45^\circ$.
This is exactly what a close scrutiny of the face-on maps of 
Englmaier \& Gerhard (1999) and Fux (1999) shows.
In addition, if the bar is itself inclined by 
$\theta_{\rm bar} \simeq 20^\circ - 25^\circ$ to the line of sight
(Englmaier \& Gerhard 1999; Fux 1999; Bissantz et al. 2003),
the $x_2$ ring/disk should be inclined by $\simeq 70^\circ$ 
to the line of sight, in excellent agreement with the observational 
conclusions of Sawada et al. (2004).
Let us also note that the gradual variation in orbit inclination with
radius automatically gives rise to spiral density enhancements.
One could surmise that these enhancements are related to the molecular arms 
discussed by Sofue (1995a) (see section~\ref{H2}).


\section{\label{s4}Our model for the gas distribution}

It is clear that our observational knowledge of the interstellar gas
distribution in the GB is rather patchy. Although theoretical models
can probably bridge some of the gaps, there remain important murky areas.
In this section, we gather the relevant observational results
(see section~\ref{s2}) that we deem the most trustworthy,
we complement them with some theoretical predictions from gas dynamical
models (see section~\ref{s3}) and we try to piece everything together 
into a coherent picture of the interstellar gas distribution 
in the innermost 3~kpc of our Galaxy.

\subsection{\label{s4_CMZ}The central molecular zone}

There have been several attempts to deproject sky maps of molecular line
emission in the CMZ. In most cases, a gas kinematic model is appealed to
in order to transform the measured line-of-sight velocity into
line-of-sight distance.
Unfortunately, our poor knowledge of the true gas kinematics 
near the GC renders this kind of method unreliable.

A notable exception is the work of Sawada et al. (2004), who derived 
a face-on map of the molecular gas without any kinematic assumption,
on the sole basis of observational (emission {\it versus} absorption) 
data (see section~\ref{H2}).
For this reason, their face-on map will serve as the first building
block of our model.
Thus, we assume that the CMZ projected onto the Galactic plane has 
the shape of a $500~{\rm pc} \times 200~{\rm pc}$ ellipse inclined 
by $70^\circ$ to the line of sight toward positive longitudes.
From Sawada et al.'s Figure~10a, we estimate that the ellipse is
centered on $(x_{\rm c},y_{\rm c}) \simeq 
(-50~{\rm pc},50~{\rm pc})$.\footnote{The $(x,y,z)$ coordinate system 
used by Sawada et al. (2004) is equivalent to our $(x,y,z)$ system 
(see Figure~\ref{fig:coordinates}) rotated clockwise by $90^\circ$.}
The peak in surface density appears displaced eastward from the ellipse
center, to $\sim (-30~{\rm pc},100~{\rm pc})$, but in view of the large
uncertainties involved in the deprojection, the apparent peak position
should not be taken at face value.
It is even possible that the molecular gas is in fact distributed
in a ring rather than a disk (e.g., Sofue 1995a). While we do not rule out 
this possibility, we find the present observational evidence inconclusive.
Theoretical studies are not of great help either. Accepting the idea
that the CMZ corresponds to the domain of $x_2$ orbits 
(see section~\ref{s3}), we note that both $x_2$ disks and $x_2$ rings 
have come up in hydrodynamical simulations 
(e.g., Jenkins \& Binney 1994; Englmaier \& Gerhard 1999; Fux 1999).

\begin{figure}
\centering
\includegraphics{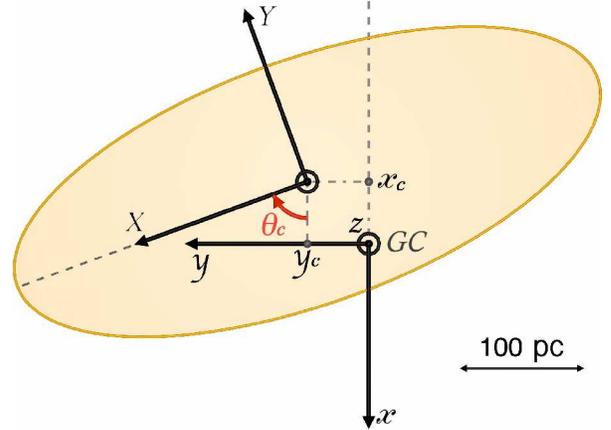} 
\caption{\label{fig:CMZ}
Face-on view of our CMZ.
$(x,y,z)$ are the Galactocentric coordinates
and $(X,Y)$ the CMZ horizontal coordinates.
The CMZ is elliptical, $500~{\rm pc} \times 200~{\rm pc}$ in size,
centered on $(x_{\rm c},y_{\rm c}) = (-50~{\rm pc},50~{\rm pc})$
and inclined by $\theta_{\rm c} = 70^\circ$ to the line of sight.
}
\end{figure}

Here, as a simple compromise, we adopt a nearly flat (i.e., between
peaked and ring-like) horizontal distribution of the type utilized 
by Launhardt et al. (2002) for their warm inner disk 
(Equation~\ref{eq:Launhardt02_disk}).
For convenience, we introduce horizontal coordinates in the CMZ frame,
$(X,Y)$, defined such that $X$ is along the major axis and $Y$ along 
the minor axis. The CMZ coordinates $(X,Y)$ are then related to 
the Galactic coordinates $(x,y)$ through
\begin{eqnarray}
\label{eq:CMZ_coord}
\left\lbrace
\begin{array}{l}
X = (x - x_{\rm c}) \ \cos \theta_{\rm c} 
  + (y - y_{\rm c}) \ \sin \theta_{\rm c} \\
Y = - (x - x_{\rm c}) \ \sin \theta_{\rm c} 
  + (y - y_{\rm c}) \ \cos \theta_{\rm c} 
\ ,
\end{array}
\right.
\end{eqnarray}
with $x_{\rm c} = -50~{\rm pc}$, $y_{\rm c} = 50~{\rm pc}$ and
$\theta_{\rm c} = 70^\circ$ (see Figure~\ref{fig:CMZ}).
Adjusting the horizontal profile from Equation~\ref{eq:Launhardt02_disk}
to an elliptical disk with semi-major axis 250~pc and axis ratio 2.5
gives for the H$_2$ space-averaged density in the CMZ
\begin{equation}
\label{eq:CMZ_r}
\langle n_{\rm H_2} \rangle _{\rm CMZ}
\ \propto \
\exp \left[ - \left( \frac{\sqrt{X^2 + (2.5 \, Y)^2} - X_{\rm c}}{L_{\rm c}} 
              \right)^4
     \right] \ ,
\end{equation}
where $X_{\rm max} = 250~{\rm pc}$, $X_{\rm c} = X_{\rm max} / 2 = 125~{\rm pc}$
and $L_{\rm c} = X_{\rm max} / (2 \, (\ln 2)^{1/4}) \simeq 137~{\rm pc}$.
Evidently, Equation~\ref{eq:CMZ_r} can easily be adapted to 
a larger/smaller CMZ or to a more/less elongated CMZ by increasing/decreasing
the value of $X_{\rm max}$ or the prefactor of $Y$, respectively.
It can also be adapted to a ring morphology by increasing 
$X_{\rm c} / X_{\rm max}$ and decreasing $L_{\rm c} / X_{\rm max}$ 
accordingly.

Sawada et al. (2004) provided only a face-on view of the molecular gas,
with no information on its vertical distribution or density scale.
Regarding the vertical distribution, one has to distinguish between 
the GC molecular disk and the surrounding 180-pc molecular ring
(see section~\ref{H2}).
The GC molecular disk appears planar and closely aligned with the
Galactic plane (Heiligman 1987; Bally et al. 1988; Burton \& Liszt 1992;
Oka et al. 1998) .
Its FWHM thickness is typically $\sim 15 - 30$~pc,
with a tendency to increase outward (up to $\sim 50$~pc) as well as 
toward massive molecular complexes (up to $\sim 30 - 60$~pc)
(Heiligman 1987; Bally et al. 1988; Sofue 1995a; Burton \& Liszt 1992).
Here, following Burton \& Liszt (1992), we ignore the spatial variations
of the disk thickness, which we set to 30~pc, and we assume a Gaussian 
vertical distribution:
\begin{equation}
\label{eq:CMZ_z}
\langle n_{\rm H_2} \rangle _{\rm CMZ}
\ \propto \
\exp \left[ - \left( \frac{z}{H_{\rm c}} \right)^2
     \right] \ ,
\end{equation}
with $H_{\rm c} = (30~{\rm pc}) / (2 \, \sqrt{\ln 2}) \simeq 18~{\rm pc}$.
The 180-pc molecular ring is thicker than the GC molecular disk
(Bally et al. 1988; Sofue 1995b), consistent with the outward increase
in the GC molecular disk's thickness noted above, 
with the flaring of the H$_2$ disk of Burton \& Liszt (1992) 
and with the greater scale height obtained by Sanders et al. (1984) 
for the H$_2$ GB disk as a whole ($r \lesssim 1.3$~kpc).
In addition, the 180-pc molecular ring is tilted out of the Galactic plane,
counterclockwise by $\sim 5^\circ - 9^\circ$ (Heiligman 1987).
It is also inclined to the line of sight, as implied by the fact that
its negative-velocity portion (near side) lies at negative latitudes,
while its positive-velocity portion (far side) lies at positive latitudes
(Heiligman 1987; Bally et al. 1988).
Heiligman (1987) estimated an inclination angle $\sim 85^\circ$ to the plane 
of the sky, whereas from Figure~11 of Bally et al. (1988) together with
the attendant discussion we estimate an inclination angle $\sim 80^\circ$.
It is noteworthy that both the tilt and the inclination of the 180-pc
molecular ring are in the same sense as those of the GB disk
(see section~\ref{H2} for the molecular GB disk and section~\ref{HI}
for the H{\sc i} GB disk).
In any event, since the 180-pc molecular ring actually contains 
but a small fraction of the CMZ mass, we will not treat it separately 
from the GC molecular disk.

We now turn to the H$_2$ density scale, which can be inferred from the total
H$_2$ mass of the CMZ. Our H$_2$ mass estimates (assuming
$X_{\rm CO} = 5 \times 10^{19}~{\rm cm^{-2}~K^{-1}~km^{-1}~s}$)
for various regions of the GB are summarized in Table~\ref{tab:MH2bis}.
Clearly, none of the listed region corresponds exactly to our CMZ,
whose projected half-size is 
$250~{\rm pc} \times \sin 70^\circ \simeq 235~{\rm pc}$.
The closest is the region $r \lesssim 200$~pc studied by Sofue (1995ab), 
which contains $\sim 1.3 \times 10^7~M_\odot$ of H$_2$ in the GC molecular disk
plus $\sim 2.7 \times 10^6~M_\odot$ of H$_2$ in the 180-pc molecular ring,
i.e., a total H$_2$ mass $\sim 1.6 \times 10^7~M_\odot$.
From this, we estimate that our CMZ has an H$_2$ mass 
$\sim 1.9 \times 10^7~M_\odot$.
This value is in good agreement with the H$_2$ mass of the GB region
$r_\perp \lesssim 375$~pc being $\sim (4.2 - 4.4) \times 10^7~M_\odot$
(Oka et al. 1998; Sodroski et al. 1995; see Table~\ref{tab:MH2bis})
and with the total hydrogen mass of the region $r \lesssim 230$~pc
being $\sim 2 \times 10^7~M_\odot$ (Launhardt et al. 2002; 
see Table~\ref{tab:MHtot}).

The H$_2$ density scale can now be obtained by integrating the density
model described by Equations~\ref{eq:CMZ_r} and \ref{eq:CMZ_z}
over the relevant region and equating the integral to
our estimated H$_2$ mass of $1.9 \times 10^7~M_\odot$.
The result is $(\langle n_{\rm H_2} \rangle _{\rm CMZ}) _{\rm max} \simeq 
150~{\rm cm}^{-3}$, 
so that the H$_2$ space-averaged density in the CMZ can be written as
\begin{eqnarray}
\label{eq:CMZ_molecular}
\langle n_{\rm H_2} \rangle _{\rm CMZ}
& \! = \! &
(150~{\rm cm}^{-3}) \,
\exp \! \left[ \! 
          - \! \left( \! 
                 \frac{\sqrt{X^2 + (2.5 \, Y)^2} - X_{\rm c}}{L_{\rm c}}
            \! \right)^4
         \right]
\nonumber \\
& & 
\times \
\exp \left[ - \left( \frac{z}{H_{\rm c}} \right)^2
     \right] \ ,
\end{eqnarray}
where $X_{\rm c} = 125~{\rm pc}$, $L_{\rm c} = 137~{\rm pc}$,
$H_{\rm c} = 18~{\rm pc}$ and $(X,Y)$ are the CMZ horizontal coordinates 
defined by Equation~\ref{eq:CMZ_coord}.
On the plane of the sky, the CMZ appears as 
a $474~{\rm pc} \times 30~{\rm pc}$ (size at half-maximum density) ellipse,
displaced eastward by 50~pc, so that it extends out to $r_\perp = 287$~pc 
at positive longitudes and $r_\perp = 187$~pc at negative longitudes
(see Figure~\ref{fig:projection_pos}a).

For the atomic gas, all the face-on maps deduced from 21-cm observations
rely on a number of kinematic assumptions, which render them questionable.
Therefore, we prefer to take an alternative approach, based on our
more solid knowledge of the molecular gas distribution and on the premise
that atomic and molecular gases are similarly distributed in space
(as argued by Liszt \& Burton 1996).
The only obvious difference between both gases concerns the thickness
of their layer, which is $\sim 30$~pc for H$_2$ (see above
Equation~\ref{eq:CMZ_z}) and $\sim 90$~pc for H{\sc i} 
(Rohlfs \& Braunsfurth 1982; see section~\ref{HI}).
Thus, we adopt for H{\sc i} the same horizontal and vertical profiles 
as for H$_2$ (Equations~\ref{eq:CMZ_r} and \ref{eq:CMZ_z},
respectively), with this difference that $H_{\rm c} \to H'_{\rm c} = 
(90~{\rm pc}) / (2 \, \sqrt{\ln 2}) \simeq 54~{\rm pc}$.

The H{\sc i} mass of the CMZ is very uncertain. As we mentioned in
section~\ref{HI}, current H{\sc i} mass estimates differ by up to 
a factor $\sim 5$.
Here, we favor the mass estimates of Rohlfs \& Braunsfurth (1982),
who directly focused on a region comparable in size to our CMZ --
as opposed to Burton \& Liszt (1978, 1993) and Liszt \& Burton (1980),
who studied the GB disk as a whole.
Rohlfs \& Braunsfurth (1982) derived an H{\sc i} mass 
$\sim 3.1 \times 10^6~M_\odot$ for the region $r \lesssim 300$~pc,
which translates into an H{\sc i} mass
$\sim 1.4 \times 10^6~M_\odot$ for the region $r \lesssim 200$~pc.
Since the H$_2$ mass in the latter region is $\sim 1.6 \times 10^7~M_\odot$
(Sofue 1995ab; see paragraph preceding Equation~\ref{eq:CMZ_molecular}), 
we may conclude that the H{\sc i} mass amounts to 8.8\% of the H$_2$ mass.

With three times the scale height and 8.8\% of the mass of the molecular gas,
the atomic gas reaches only 2.9\% of its maximum mass density, 
or 5.8\% of its maximum number density, i.e.,
$(\langle n_{\rm H{\scriptscriptstyle I}} \rangle _{\rm CMZ}) _{\rm max} 
\simeq 8.8~{\rm cm}^{-3}$.
Altogether, the H{\sc i} space-averaged density in the CMZ is given by
\begin{eqnarray}
\label{eq:CMZ_atomic}
\langle n_{\rm H{\scriptscriptstyle I}} \rangle _{\rm CMZ}
& \! = \! &
(8.8~{\rm cm}^{-3}) \,
\exp \! \left[ \! 
          - \! \left( \! 
                 \frac{\sqrt{X^2 + (2.5 \, Y)^2} - X_{\rm c}}{L_{\rm c}}
            \! \right)^4
         \right]
\nonumber \\
& & 
\times \
\exp \left[ - \left( \frac{z}{H'_{\rm c}} \right)^2
     \right] \ ,
\end{eqnarray}
with $H'_{\rm c} = 54~{\rm pc}$ and $X_{\rm c}$, $L_{\rm c}$, $(X,Y)$ 
as defined below Equation~\ref{eq:CMZ_molecular}
(see Figure~\ref{fig:projection_pos}b).

\subsection{\label{s4_disk}The ``holed" GB disk}

We saw in section~\ref{s4_CMZ} that reliable observational information
on the interstellar gas distribution inside the CMZ is scanty. It turns out 
that the observational situation is even worse outside the CMZ.
It is true that a number of tilted GB disk models have been constructed
on the basis of CO and 21-cm observations (see sections~\ref{H2} and 
\ref{HI}), but all of them involve unverified assumptions, both on 
the overall geometry (e.g., axial symmetry) and on the gas kinematics.

The model that we consider the most plausible, despite its inherent
uncertainties, is that of Liszt \& Burton (1980), because it is the only
model with an elongated disk, which makes it the easiest to reconcile 
with our theoretical understanding of gas dynamics in the GB 
(see section~\ref{s3}).
Indeed, the GB disk outside the CMZ is believed to correspond to the domain
of non self-intersecting $x_1$ orbits, which form an elongated, radially
thick ring around the $x_2$ disk/ring (itself identified with the CMZ).
In some hydrodynamical simulations, the inner boundary of the $x_1$ ring 
grazes the $x_2$ disk/ring, while in others, a gap opens up between both.

Here, we start from Liszt \& Burton's (1980) GB disk, namely, 
an elliptical disk with semi-major axis 1.6~kpc and axis ratio 3.1,
and we introduce a hole in the middle just large enough to enclose the CMZ.
It can be checked that a hole with the same shape and half the size
of the entire GB disk is sufficient for our purpose.
Then the inner boundary of the ``holed" GB disk has a half-size of
$800~{\rm pc} \times 258~{\rm pc}$, in reasonably good agreement with 
the results of hydrodynamical simulations.
For instance, Jenkins \& Binney (1994) found that the inner boundary of
the $x_1$ ring has a half-size
$\simeq (750 \pm 50)~{\rm pc} \times (400 \pm 50)~{\rm pc}$.
The $x_1$ ring is more difficult to make out in the larger-scale
simulated maps of Englmaier \& Gerhard (1999) and of Bissantz et al. (2003),
but its inner boundary clearly appears more elongated, with an axis ratio 
possibly close to our adopted value of 3.1.
Let us also point out that our GB disk is more elongated than our CMZ
(whose axis ratio is 2.5; see section~\ref{s4_CMZ}). This is perfectly
consistent not only with gas flow simulations, but also with particle
orbit calculations, which predict a greater ellipticity for $x_1$ orbits 
than for $x_2$ orbits (see first paragraph of section~\ref{s3}).

\begin{figure}
\centering
\includegraphics{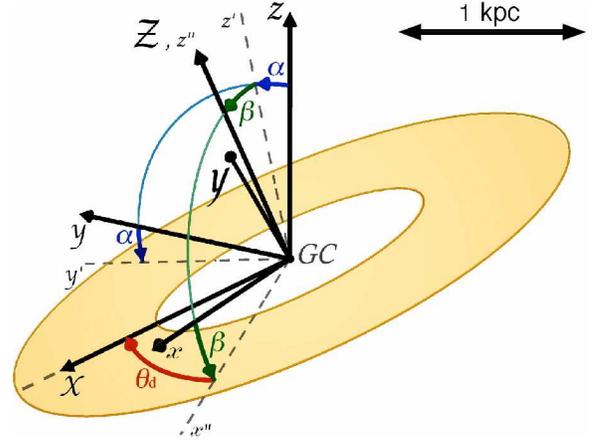} 
\caption{\label{fig:disk}
Three-dimensional view of our holed GB disk.
$(x,y,z)$ are the Galactocentric coordinates,
$({\cal X}, {\cal Y}, {\cal Z})$ the GB disk coordinates,
and $(x,y',z')$, $(x'',y',z'')$ the auxiliary coordinates
defined above Equation~\ref{eq:disk_coord}.
The holed GB disk is elliptical, 
3.2~kpc long and $(3.2~{\rm kpc}) / 3.1 \simeq 1.03~{\rm kpc}$ wide,
and its inner boundary has half the dimensions of its outer boundary.
Relative to the Galactic plane, its principal plane 
is tilted by $\alpha = 13\fdg5$ about the $x$-axis
and inclined by $\beta = 20^\circ$ about the $y'$-axis,
and its major axis is oriented at $\theta_{\rm d} = 48\fdg5$ 
to the $x''$-axis.
}
\end{figure}

According to Liszt \& Burton (1980), the GB disk is tilted by 
$\alpha = 13\fdg5$ out of the Galactic plane [counterclockwise rotation
by $\alpha$ about the $x$-axis, whereby $(x,y,z) \to (x,y',z')$]
and inclined by $i = 70^\circ$ to the plane of the sky
[``front-downward" rotation by $\beta = 90^\circ - i$ about the $y'$-axis, 
whereby $(x,y',z') \to (x'',y',z'')$], and the GB disk's major axis
forms an angle $\theta_{\rm d} = 48\fdg5$ to the $x''$-axis
[measured clockwise about the $z''$-axis].
If we denote the spatial coordinates in the GB disk frame by
$({\cal X}, {\cal Y}, {\cal Z})$, with ${\cal X}$ along the major axis,
${\cal Y}$ along the minor axis and ${\cal Z}$ along the northern normal,
the transformation from Galactic coordinates to GB disk coordinates reads
\begin{eqnarray}
\label{eq:disk_coord}
\left\lbrace
\begin{array}{lll}
{\cal X} & = & x \ \cos \beta \ \cos \theta_{\rm d}
\\
& & - \ y \ \left( \sin \alpha \ \sin \beta \ \cos \theta_{\rm d}
                 - \cos \alpha \ \sin \theta_{\rm d}
          \right)
\\
& & - \ z \ \left( \cos \alpha \ \sin \beta \ \cos \theta_{\rm d}
                 + \sin \alpha \ \sin \theta_{\rm d}
          \right)
\\
{\cal Y} & = & - x \ \cos \beta \ \sin \theta_{\rm d}
\\
& & + \ y \ \left( \sin \alpha \ \sin \beta \ \sin \theta_{\rm d}
                 + \cos \alpha \ \cos \theta_{\rm d}
          \right)
\\
& & + \ z \ \left( \cos \alpha \ \sin \beta \ \sin \theta_{\rm d}
                 - \sin \alpha \ \cos \theta_{\rm d}
          \right)
\\
{\cal Z} & = & x \ \sin \beta 
\\
& & + \ y \ \sin \alpha \ \cos \beta
\\
& & + \ z \ \cos \alpha \ \cos \beta
\ ,
\end{array}
\right.
\end{eqnarray}
with $\alpha = 13\fdg5$, $\beta = 20^\circ$ and $\theta_{\rm d} = 48\fdg5$
(see Figure~\ref{fig:disk}).

Now that we have specified the horizontal shape and dimensions as well as
the orientation of our holed GB disk, we may fit to it horizontal and 
vertical profiles of the same type as for our CMZ (Equations~\ref{eq:CMZ_r} 
and \ref{eq:CMZ_z}, respectively). Again, we assume that molecular and atomic 
gases are similarly distributed in space, except for a possible difference 
in scale height. Under these conditions, we can write
\begin{eqnarray*}
\label{eq:disk_r}
\langle n_{\rm H_2} \rangle _{\rm disk}
& \propto &
\exp \left[ - \left( 
                \frac{\sqrt{{\cal X}^2 + (3.1 \, {\cal Y})^2} 
                      - {\cal X}_{\rm d}}
                     {L_{\rm d}} 
              \right)^4
     \right] \ ,
\nonumber \\
\langle n_{\rm H{\scriptscriptstyle I}} \rangle _{\rm disk}
& \propto &
\exp \left[ - \left( 
                \frac{\sqrt{{\cal X}^2 + (3.1 \, {\cal Y})^2} 
                      - {\cal X}_{\rm d}}
                     {L_{\rm d}} 
              \right)^4
     \right] \ ,
\end{eqnarray*}
\vspace*{-1.8cm}

\rightline{(\ref{eq:disk_r})}
\addtocounter{equation}{1}
\vspace*{-12pt}
\vspace*{1.8cm}

\noindent
with ${\cal X}_{\rm max} = 1.6~{\rm kpc}$, 
${\cal X}_{\rm min} = 0.8~{\rm kpc}$,
${\cal X}_{\rm d} = ({\cal X}_{\rm max} + {\cal X}_{\rm min}) / 2 
= 1.2~{\rm kpc}$ and 
$L_{\rm d} = ({\cal X}_{\rm max} - {\cal X}_{\rm min}) / (2 \, (\ln 2)^{1/4}) 
\simeq 438~{\rm pc}$, and
\begin{eqnarray*}
\label{eq:disk_z}
\langle n_{\rm H_2} \rangle _{\rm disk}
& \propto &
\exp \left[ - \left( \frac{{\cal Z}}{H_{\rm d}} \right)^2
     \right] \ ,
\nonumber \\
\langle n_{\rm H{\scriptscriptstyle I}} \rangle _{\rm disk}
& \propto &
\exp \left[ - \left( \frac{{\cal Z}}{H'_{\rm d}} \right)^2
     \right] \ \cdot
\end{eqnarray*}
\vspace*{-1.6cm}

\rightline{(\ref{eq:disk_z})}
\addtocounter{equation}{1}
\vspace*{-12pt}
\vspace*{1.6cm}

\noindent
For the scale height, we adopt
$H_{\rm d} = (70~{\rm pc}) / (2 \, \sqrt{\ln 2}) \simeq 42~{\rm pc}$
for H$_2$ (Sanders et al. 1984)
and $H'_{\rm d} = \sqrt{2} \ (85~{\rm pc}) \simeq 120~{\rm pc}$ 
for H{\sc i} (Liszt \& Burton 1980).
Like in the CMZ, the atomic layer is about three times thicker than
the molecular layer. Moreover, each layer is $\simeq 2.2-2.3$ times thicker 
in the holed GB disk than in the CMZ.

Our density model can again be normalized with the help of
the total H$_2$ and H{\sc i} masses in the considered region.
The entire GB disk contains $\sim 5.3 \times 10^7~M_\odot$ of H$_2$ 
(Sanders et al. 1984; see Table~\ref{tab:MH2bis})
and $\sim 5.2 \times 10^6~M_\odot$ of H{\sc i} 
(Liszt \& Burton 1980; see Table~\ref{tab:MHI}),
while our CMZ contains $\sim 1.9 \times 10^7~M_\odot$ of H$_2$
and $\sim 1.7 \times 10^6~M_\odot$ of H{\sc i} 
(see section~\ref{s4_CMZ}).
By subtraction, we find that our holed GB disk has 
an H$_2$ mass $\sim 3.4 \times 10^7~M_\odot$ and 
an H{\sc i} mass $\sim 3.5 \times 10^6~M_\odot$.
We are fully aware of the pitfalls of this mass estimation,
especially as the masses involved in the subtraction procedure 
have considerable uncertainty.
Particularly uncertain is the H$_2$ mass of the entire GB disk.
The only estimate available to us is the value from Sanders et al. (1984).
But this value applies to an axisymmetric disk of radius 1.3~kpc
having twice the surface area of our elliptical GB disk, which means that
it could overestimate the true mass by up to a factor $\sim 2$.
On the other hand, the comparatively large values obtained by
Sodroski et al. (1995) and by Oka et al. (1998) for a smaller region
(see Table~\ref{tab:MH2bis}) suggest instead that Sanders et al.'s (1984) 
value could underestimate the true mass by roughly the same factor.
Hence, the two sources of error tend to counteract each other.
In addition, two facts allow us to place some confidence in our results.
First, the H$_2$ and H{\sc i} masses of our holed GB disk are slightly 
smaller than what they would be if the gas were uniformly distributed 
throughout the GB disk ($\sim 4.0 \times 10^7~M_\odot$ and
$\sim 3.9 \times 10^6~M_\odot$, respectively).
Second, H{\sc i} represents 8.8\% by mass of H$_2$ in the CMZ
and 10.3\% in the holed GB disk. This statement supports Liszt \& Burton's
(1996) claim that both gases have similar spatial distributions,
not only in the CMZ and in the holed GB disk separately 
(as we already assumed earlier), but also throughout the entire GB disk.
It is also reassuring to obtain a slightly lower H{\sc i} mass fraction 
in the CMZ.

\begin{figure}
\centering
\includegraphics{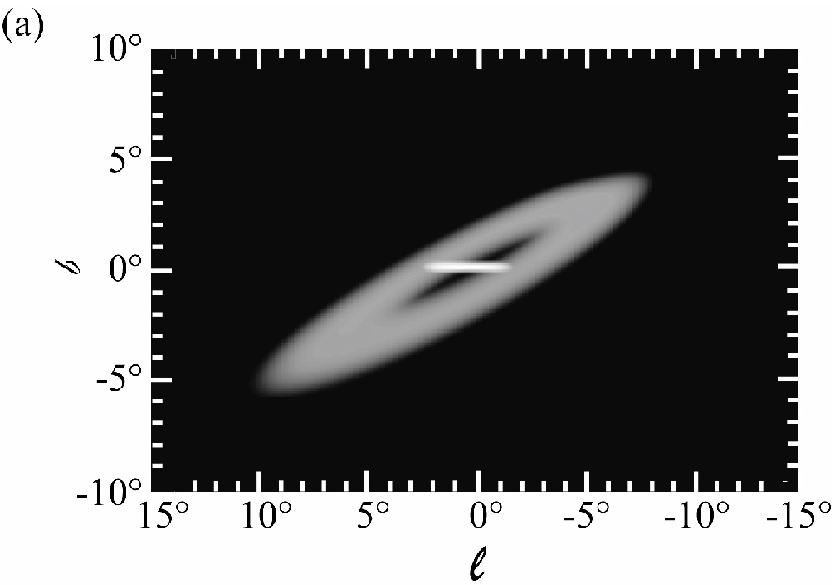}
\bigskip\bigskip\\
\includegraphics{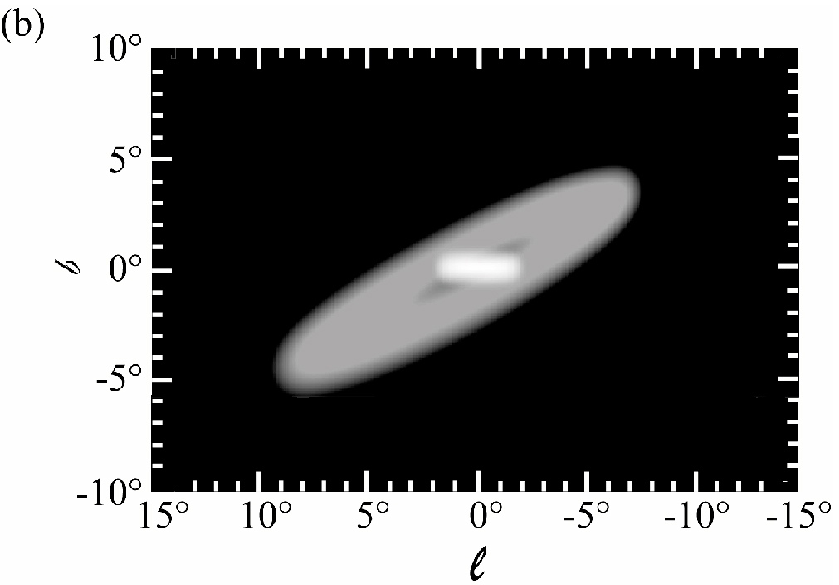}
\caption{\label{fig:projection_pos}
Projection of the CMZ (bright area) and the holed GB disk (fainter area) 
onto the plane of the sky: (a) molecular gas 
(see Equations~\ref{eq:CMZ_molecular} and \ref{eq:disk_molecular});
(b) atomic gas (see Equations~\ref{eq:CMZ_atomic} and \ref{eq:disk_atomic}).
The apparent sizes are a little larger than the sizes at half-maximum
density, because of the logarithmic scale used in the projection.
In contrast to the CMZ, which is truly displaced to the left, 
the GB disk is symmetric with respect to the GC, and the only reason why 
it appears more extended on the left side is because its positive-longitude 
portion lies closer to us.
}
\end{figure}

\begin{figure}
\centering
\includegraphics{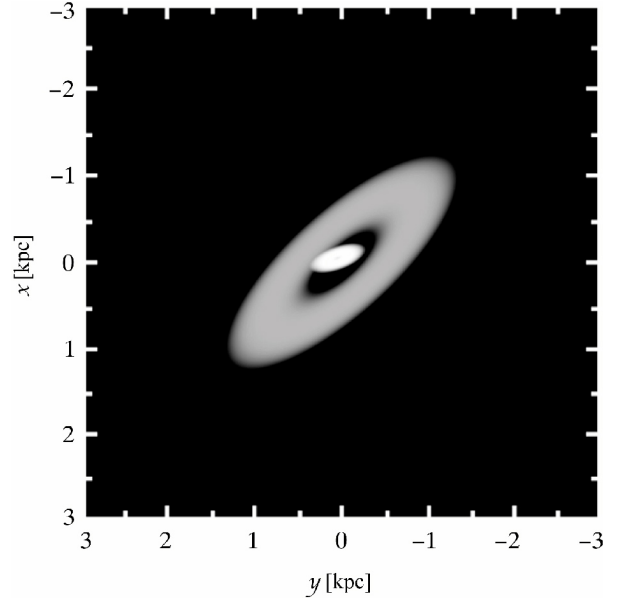}
\caption{\label{fig:projection_GP}
Projection of the CMZ (bright area) and the holed GB disk (fainter area) 
onto the Galactic plane.
Displayed here is the H$_2$ map 
(from Equations~\ref{eq:CMZ_molecular} and \ref{eq:disk_molecular}).
The H{\sc i} map
(from  Equations~\ref{eq:CMZ_atomic} and \ref{eq:disk_atomic})
looks identical, except for this hardly noticeable difference that
the GB-disk--to--CMZ luminosity ratio is slightly greater.
For the same reason as in Figure~\ref{fig:projection_pos}, the apparent sizes 
are a little larger than the sizes at half-maximum density.
}
\end{figure}

If we now equate the space integral of Equation~\ref{eq:disk_r} times
Equation~\ref{eq:disk_z} to our estimated H$_2$ and H{\sc i} masses,
we find 
$(\langle n_{\rm H_2} \rangle _{\rm disk}) _{\rm max} \simeq 4.8~{\rm cm}^{-3}$
and $(\langle n_{\rm H{\scriptscriptstyle I}} \rangle _{\rm disk}) _{\rm max}
\simeq 0.34~{\rm cm}^{-3}$.
Hence, the holed GB disk has an H$_2$ space-averaged density 
\begin{eqnarray}
\label{eq:disk_molecular}
\langle n_{\rm H_2} \rangle _{\rm disk}
& \! = \! &
(4.8~{\rm cm}^{-3}) \,
\exp \! \left[ \! 
          - \! \left( \! 
                 \frac{\sqrt{{\cal X}^2 + (3.1 \, {\cal Y})^2} 
                       - {\cal X}_{\rm d}}
                      {L_{\rm d}} 
            \! \right)^4
         \right]
\nonumber \\
& & 
\times \
\exp \left[ - \left( \frac{{\cal Z}}{H_{\rm d}} \right)^2
     \right]
\end{eqnarray}
and an H{\sc i} space-averaged density
\begin{eqnarray}
\label{eq:disk_atomic}
\langle n_{\rm H{\scriptscriptstyle I}} \rangle _{\rm disk}
& \! = \! &
(0.34~{\rm cm}^{-3}) \,
\exp \! \left[ \! 
          - \! \left( \! 
                 \frac{\sqrt{{\cal X}^2 + (3.1 \, {\cal Y})^2} 
                       - {\cal X}_{\rm d}}
                      {L_{\rm d}} 
            \! \right)^4
         \right]
\nonumber \\
& & 
\times \
\exp \left[ - \left( \frac{{\cal Z}}{H'_{\rm d}} \right)^2
     \right] \ ,
\end{eqnarray}
with ${\cal X}_{\rm d} = 1.2~{\rm kpc}$, $L_{\rm d} = 438~{\rm pc}$, 
$H_{\rm d} = 42~{\rm pc}$ and $H'_{\rm d} = 120~{\rm pc}$.
On the plane of the sky, the GB disk extends out to $r_\perp = 1.14$~kpc
(radius at half-maximum density) on each side of the GC
(see Figure~\ref{fig:projection_pos}). 
Projected onto the Galactic plane, it has the shape of 
a $2.94~{\rm kpc} \times 1.02~{\rm kpc}$ (FWHM size) ellipse inclined 
clockwise by $47\fdg6$ to the line of sight 
(see Figure~\ref{fig:projection_GP}).
This inclination angle is greater than that typically found for
the Galactic stellar bar ($\theta_{\rm bar} \simeq 15^\circ - 35^\circ$;
see section~\ref{s3}), but it is in good agreement with the value 
$\theta_{\rm bar} = 44^\circ \pm 10^\circ$ recently obtained
by Benjamin et al. (2005) from the GLIMPSE Point Source Catalog.

\subsection{\label{s4_ionized}The ionized component}

The best available model for the spatial distribution of interstellar
free electrons in the GB is the NE2001 model of Cordes \& Lazio (2002)
presented in section~\ref{HII}. According to this model, the total mass 
of interstellar ionized hydrogen in the region $r \le 3$~kpc is 
$(7.3 \times 10^7~M_\odot) / (1 + 0.2 \ f_{\rm HIM})$, 
where $f_{\rm HIM}$ is the fraction of ionized gas belonging to the hot
medium (see Table~\ref{tab:MHII}).
The mass of hot H$^+$ in the same region can be estimated from Almy et al.'s 
(2000) model (neglecting the contribution from very hot H$^+$) at
$1.2 \times 10^7~M_\odot$ (see Table~\ref{tab:MHII}).
It then follows that $f_{\rm HIM} = 17\%$ 
(or, equivalently, $f_{\rm WIM} = 83\%$)
and that the total mass of H$^+$ inside 3~kpc is $7.1 \times 10^7~M_\odot$,
divided between $5.9 \times 10^7~M_\odot$ in the WIM and
$1.2 \times 10^7~M_\odot$ in the HIM.
Furthermore, from Equation~\ref{eq:n_e} with $f_{\rm HIM} = 17\%$,
we gather that the H$^+$ space-averaged density is given by 
$\langle n_{\rm H^+} \rangle = 0.97 \ \langle n_{\rm e} \rangle$.
The partial contributions from the warm and hot ionized media 
are globally given by $\langle n_{\rm H^+} \rangle _{\rm _{WIM}} = 
f_{\rm WIM} \ \langle n_{\rm H^+} \rangle$
and $\langle n_{\rm H^+} \rangle _{\rm _{HIM}} = 
f_{\rm HIM} \ \langle n_{\rm H^+} \rangle$, respectively.
For the WIM, which contributes a large 83\% of the total H$^+$ mass, 
we may reasonably assume that the above global relation remains 
approximately valid locally.
Owing to the large uncertainties in the exact spatial dependence of
the density distributions, we feel that taking 
$\langle n_{\rm H^+} \rangle _{\rm _{WIM}} = 
f_{\rm WIM} \ \langle n_{\rm H^+} \rangle$ at all ${\bf r}$ 
is safer than subtracting $\langle n_{\rm H^+} \rangle _{\rm _{HIM}}$
(which can be estimated independently; see next paragraph) from 
$\langle n_{\rm H^+} \rangle$.
In that case, the H$^+$ space-averaged density of the WIM is simply
$\langle n_{\rm H^+} \rangle _{\rm _{WIM}} = 0.80 \ \langle n_{\rm e} \rangle$
or, in view of Equations~\ref{eq:Cordes05} -- \ref{eq:Cordes05_GC},
\begin{eqnarray*}
\label{eq:ionized_WIM}
\langle n_{\rm H^+} \rangle _{\rm _{WIM}}
& = & 
(8.0~{\rm cm}^{-3}) \
\nonumber \\
& \hspace*{-3cm} \times &
\hspace*{-1.5cm} 
\left\{
\exp \left[ - \frac{x^2 + (y - y_3)^2}{L_3^2} 
     \right] \
\exp \left[ - \frac{(z - z_3)^2}{H_3^2}
     \right]
\right.
\nonumber \\
& &
\hspace*{-1.5cm} 
\ + \ 0.009 \ 
\exp \left[ - \left( \frac{r - L_2}{L_2/2} \right)^2
     \right] \
{\rm sech}^2 \left( \frac{z}{H_2} \right)
\nonumber \\
& &
\left.
\hspace*{-1.5cm} 
\ + \ 0.005 \ 
\left[ \cos \left( \pi \ \frac{r}{2 \, L_1} \right) \
       u ( L_1 - r )
\right] \
{\rm sech}^2 \left( \frac{z}{H_1} \right)
\right\} \ ,
\end{eqnarray*}
\rightline{(\ref{eq:ionized_WIM})}
\addtocounter{equation}{1}
\vspace*{-8pt}

\noindent
with $y_3 = -10$~pc, $z_3 = -20$~pc, $L_3 = 145$~pc, $H_3 = 26$~pc,
$L_2 = 3.7$~kpc, $H_2 = 140$~pc, $L_1 = 17$~kpc and $H_1 = 950$~pc.

For the HIM, we turn again to Almy et al.'s (2000) hydrostatic-polytropic 
model. With $\langle n_{\rm H^+} \rangle _{\rm _{HIM}} = 
\langle n_{\rm e} \rangle _{\rm _{HIM}} / 1.2$,
the H$^+$ space-averaged density of the HIM directly follows from
Equation~\ref{eq:Almy00}:
\begin{eqnarray}
\label{eq:ionized_HIM}
\langle n_{\rm H^+} \rangle _{\rm _{HIM}}
& = &
\Big\{ 
(0.009~{\rm cm}^{-3})^{2/3}
- (1.54 \times 10^{-17}~{\rm cm}^{-4}~{\rm s}^2)
\nonumber \\
& &
\quad \times \
\big[ \phi(r,z) - \phi(0,0) \big]
\Big\} ^{1.5} \ ,
\end{eqnarray}
where $\phi(r,z)$ is the Galactic gravitational potential 
of Wolfire et al. (1995) given by Equation~\ref{eq:Wolfire95}.

For completeness, we also provide a description of the very hot medium 
detected by Koyama et al. (1989) and discussed at the end 
of section~\ref{HII}.
Based on Yamauchi et al.'s (1990) and Koyama et al.'s (1996) studies,
we assign to the VHIM an ellipsoid Gaussian density distribution with 
central electron density $0.35~{\rm cm}^{-3}$ and
FWHM size $(270~{\rm pc})^2 \times 150~{\rm pc}$.
The major plane of the ellipsoid is seen edge-on from the Sun
and tilted clockwise by $21^\circ$ with respect to the Galactic plane, 
which leads us to introduce the coordinate transformation
\begin{eqnarray}
\label{eq:VHIM_coord}
\left\lbrace
\begin{array}{l}
\eta = y \ \cos \alpha_{\rm vh} + z \ \sin \alpha_{\rm vh} \\
\zeta = - y \ \sin \alpha_{\rm vh} + z \ \cos \alpha_{\rm vh} 
\ ,
\end{array}
\right.
\end{eqnarray}
with $\alpha_{\rm vh} = 21^\circ$.
The H$^+$ space-averaged density of the VHIM,
$\langle n_{\rm H^+} \rangle _{\rm _{VHIM}} = 
\langle n_{\rm e} \rangle _{\rm _{VHIM}} / 1.2$,
can then be written in terms of the $(\eta,\zeta)$ coordinates as
\begin{equation}
\label{eq:ionized_VHIM}
\langle n_{\rm H^+} \rangle _{\rm _{VHIM}}
\ = \
(0.29~{\rm cm}^{-3}) \ 
\exp \left[ - \left( \frac{x^2 + \eta^2}{L_{\rm vh}^2} 
                     + \frac{\zeta^2}{H_{\rm vh}^2}
              \right)
     \right] \ ,
\end{equation}
with $L_{\rm vh} = 162~{\rm pc}$ and $H_{\rm vh} = 90~{\rm pc}$.

\subsection{\label{s4_together}Altogether}

The total space-averaged density of hydrogen nuclei in the interstellar GB
is given by the sum of the partial contributions from the molecular, atomic 
and ionized media:
\begin{equation}
\label{eq:density_total}
\langle n_{\rm H} \rangle
\ = \
2 \ \langle n_{\rm H_2} \rangle
\ + \ \langle n_{\rm H{\scriptscriptstyle I}} \rangle
\ + \ \langle n_{\rm H^+} \rangle \ ,
\end{equation}
where 
\begin{equation}
\label{eq:density_molecular}
\langle n_{\rm H_2} \rangle
\ = \
\langle n_{\rm H_2} \rangle _{\rm CMZ}
\ + \ \langle n_{\rm H_2} \rangle _{\rm disk}
\end{equation}
(see Equations~\ref{eq:CMZ_molecular} and \ref{eq:disk_molecular}),
\begin{equation}
\label{eq:density_atomic}
\langle n_{\rm H{\scriptscriptstyle I}} \rangle
\ = \
\langle n_{\rm H{\scriptscriptstyle I}} \rangle _{\rm CMZ}
\ + \ \langle n_{\rm H{\scriptscriptstyle I}} \rangle _{\rm disk}
\end{equation}
(see Equations~\ref{eq:CMZ_atomic} and \ref{eq:disk_atomic}) and
\begin{equation}
\label{eq:density_ionized}
\langle n_{\rm H^+} \rangle
\ = \
\langle n_{\rm H^+} \rangle _{\rm _{WIM}}
\ + \ \langle n_{\rm H^+} \rangle _{\rm _{HIM}}
\ + \ \langle n_{\rm H^+} \rangle _{\rm _{VHIM}}
\end{equation}
(see Equations~\ref{eq:ionized_WIM}, \ref{eq:ionized_HIM} and
\ref{eq:ionized_VHIM}).
For an assumed total-to-hydrogen mass ratio of 1.453 (see beginning of 
section~\ref{s2}), the total space-averaged mass density of interstellar gas 
in the interstellar GB is related to the above hydrogen space-averaged 
density through
\begin{equation}
\label{eq:mass_density}
\langle \rho \rangle =
1.453 \ m_{\rm P} \ \langle n_{\rm H} \rangle \ ,
\end{equation}
where $m_{\rm P}$ is the proton rest mass.

\begin{figure}
\centering
\includegraphics{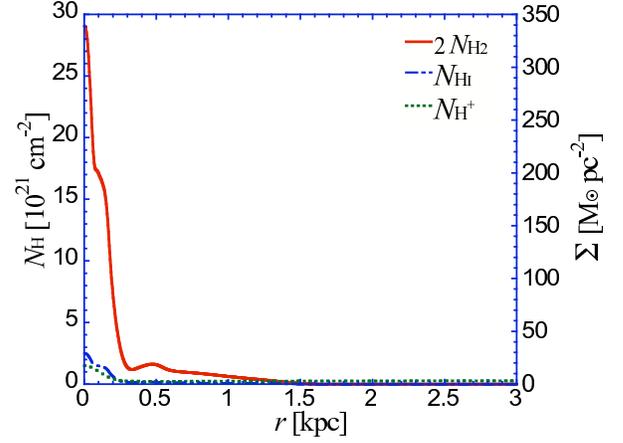}
\caption{\label{fig:density_hor}
Azimuthally-averaged column densities through the Galactic disk
of interstellar hydrogen nuclei in molecular form (solid line), 
atomic form (dot-dashed line) and ionized form (dotted line) 
and associated surface densities of total interstellar matter 
(assuming a total-to-hydrogen mass ratio of 1.453; see section~\ref{s2})
as functions of Galactic radius.
}
\end{figure}

\begin{figure}
\centering
\includegraphics{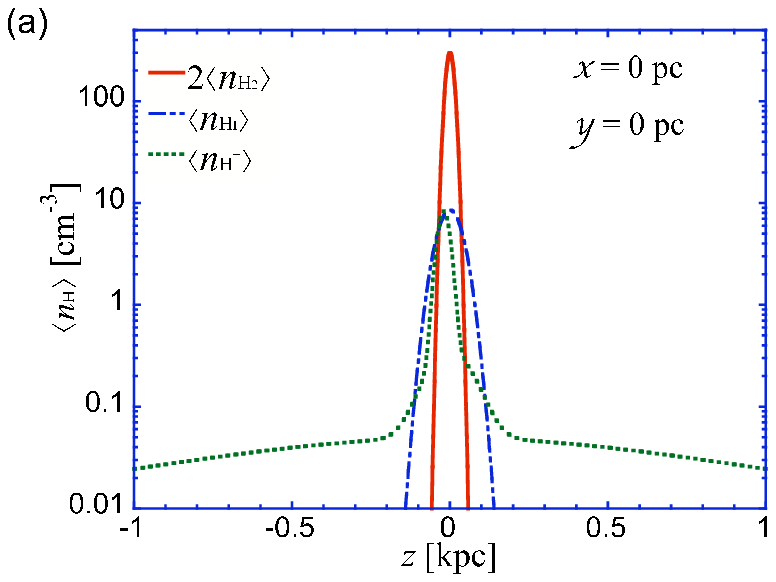}
\bigskip\bigskip\\
\includegraphics{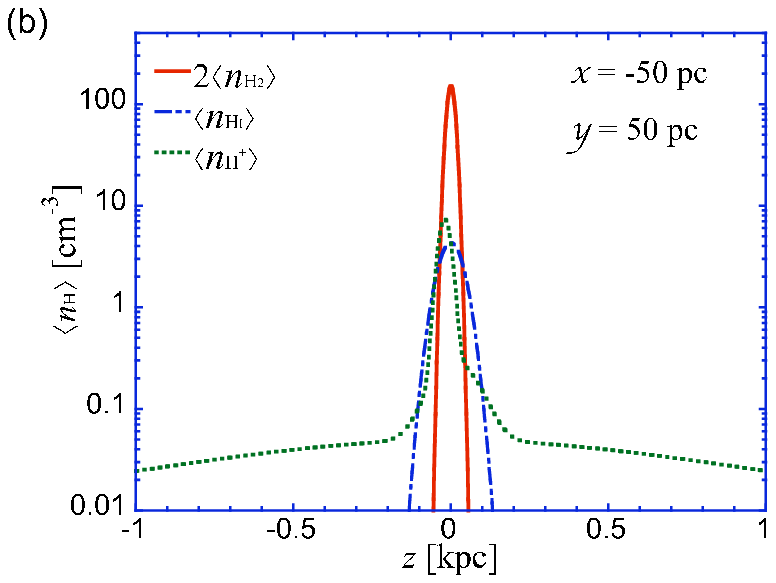}
\bigskip\bigskip\\
\includegraphics{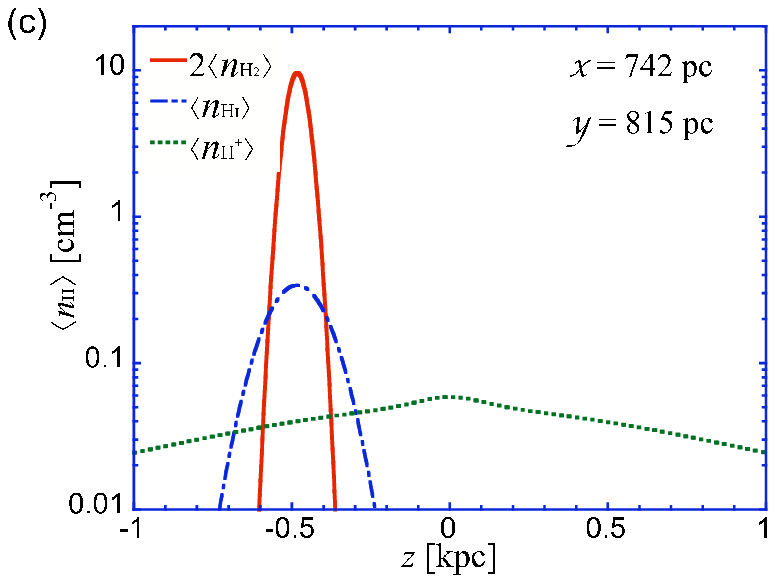}
\caption{\label{fig:density_vert}
Space-averaged densities of interstellar hydrogen nuclei 
in molecular form (solid line), atomic form (dot-dashed line)
and ionized form (dotted line) as functions of Galactic
height: (a) at the GC; (b) at the center of the CMZ;
(c) at the point of the GB disk (projected onto the Galactic plane) 
where its major axis intersects its ``ridge".
}
\end{figure}

Displayed in Figure~\ref{fig:density_hor} is the radial variation
of the azimuthally-averaged column densities through the Galactic disk
of the molecular, atomic and ionized gases.
The vertical variation of their hydrogen space-averaged densities
is shown in Figure~\ref{fig:density_vert}
at three representative horizontal locations, namely, 
at the GC [$(x,y) = (0,0)$], 
at the center of the CMZ [$(x,y) = (-50~{\rm pc},50~{\rm pc})$]
and at the point of the GB disk (projected onto the Galactic plane) 
where its major axis intersects its ``ridge"
[$r = \frac{3}{4} \, (1.47~{\rm kpc})$, $\theta = 47\fdg6$
(see below Equation~\ref{eq:disk_atomic}), i.e., 
$(x,y) = (742~{\rm pc},815~{\rm pc})$].
The graphs speak for themselves.
The molecular gas (solid line) is by far the most abundant.
Its radial distribution exhibits a high peak at the origin 
corresponding to the CMZ and a lower bump at $r \simeq 4.7$~kpc 
corresponding to the holed GB disk.
Along the vertical, it is confined to a thin layer, which
is nearly centered on the Galactic midplane,
except outside the CMZ, where it is centered on the tilted midplane
of the GB disk (see Figure~\ref{fig:density_vert}c).
The atomic gas (dot-dashed line) is about ten times less abundant
than the molecular gas, it has virtually the same radial distribution 
and it occupies an approximately three times thicker layer.
Finally, the ionized gas (dotted line) is the most rarefied,
and it spreads much farther out in all directions.
The slight rise in its radial distribution outside $\sim 1$~kpc
is due to the contribution from the Galactic disk.

\section{\label{s5}Conclusions}

In this paper, we took stock of the present observational status 
of interstellar gas in the Galactic bulge (defined as the region
interior to $r = 3$~kpc) and we discussed in some detail existing
theoretical models of interstellar gas dynamics near the GC.
We then provided simple analytical expressions for the hydrogen
space-averaged densities of the different gas components that fit at best 
the observational data and are consistent with the theoretical predictions.
All our expressions are written in parametric form, which will make them
easy to update and adapt to future observations showing other 
characteristics (shape, dimensions, orientation, mass) than those 
adopted here. Our results are summarized in section~\ref{s4_together}.

We did not address the issue of where our modeled interstellar gas 
would appear in an $(l,v)$ or full $(l,b,v)$ plot.
Doing so would require studying the gas dynamics in a specified
non-axisymmetric Galactic gravitational potential with a given pattern speed.
This is undoubtedly a step that should be taken in the future,
especially as the $(l,b,v)$ space is the one that observers have direct
access to.

Aside from the space-averaged densities, we were able to gather
some interesting observational information on the physical conditions
(temperature and true density) in the molecular and ionized media.
In section~\ref{H2}, we saw that the molecular medium probably contains
a cold, high-density component with $T \sim 50$~K and
$n_{{\rm H}_2} \sim 10^{3.5-4}~{\rm cm}^{-3}$,
and a warm, low-density component with $T \sim 150$~K and
$n_{{\rm H}_2} \sim 10^{2.5}~{\rm cm}^{-3}$.
It is also possible that the temperature and true density in 
the molecular medium vary more-or-less continuously over broad ranges
of values.
In section~\ref{HII}, we were led to describe the ionized medium as 
the superposition of three components with radically different temperatures:
the WIM with $T \sim 10^4$~K, the HIM with $T \sim$ a few $10^6$~K
and the VHIM with $T \gtrsim 10^8$~K. However, we were unable to obtain 
independent estimates for the associated true densities.
As for the atomic medium, we may conjecture by analogy with
the interstellar vicinity of the Sun that a cold and a warm phase co-exist, 
although we found little direct evidence to substantiate this hypothesis
and even less direct quantitative information on the physical conditions
in these two putative phases.

It would be highly desirable to estimate the filling factors of 
the different media. Unfortunately, this cannot be done in any
meaningful way for the atomic and ionized media, whose true densities
are not known.
In contrast, crude upper limits can be derived for the two components 
of the molecular medium.
Remember that the filling factor is by definition the ratio of 
space-averaged to true density.
The H$_2$ space-averaged density attains its maximum value in the CMZ,
where $\langle n_{\rm H_2} \rangle _{\rm max} = 150~{\rm cm}^{-3}$
(see Equation~\ref{eq:CMZ_molecular}).
If, as argued by Rodr\'iguez-Fern\'andez et al. (2001), $\sim 30\%$ 
of the interstellar molecular gas resides in the warm component 
(see section~\ref{H2}) and if the cold and warm components have 
true densities $\sim 10^{3.5-4}~{\rm cm}^{-3}$ and
$\sim 10^{2.5}~{\rm cm}^{-3}$, respectively, then their filling factors 
are $\lesssim 3.3\%$ and $\lesssim 14\%$, respectively.

\begin{acknowledgement}{}
The authors would like to thank W.B.~Burton, J.~Cordes, F.~Combes, T.~Dame, 
R.~Fux, R.~Launhardt, J.~Lazio, D.~McCammon, C.~Martin, N.~McClure-Griffiths,
N.~Rodriguez, and the referee, J. Binney, for their helpful comments.
\end{acknowledgement}




\begin{thebibliography}{}
\bibitem[Afflerbach et al. 1997]{}
Afflerbach, A., Churchwell, E., \& Werner, M. W. 1997, \apj~478, 190

\bibitem[Almy et al. 2000]{}
Almy, R. C., McCammon, D., Digel, S. W., et al. 2000, \apj~545, 290

\bibitem[Anders and Grevesse 1988]{}
Anders, A., \& Grevesse, N. 1988, Geochimica et Cosmochimica Acta~53, 197
	
\bibitem[Arimoto et al. 1996]{}
Arimoto, N., Sofue, Y., \& Tsujimoto, T. 1996, PASJ~48, 275

\bibitem[Athanassoula 1992a]{}
Athanassoula, E. 1992, \mnras~259, 328

\bibitem[Athanassoula 1992b]{}
Athanassoula, E. 1992, \mnras~259, 345

\bibitem[Bally et al. 1987]{}
Bally, J., Stark, A. A., Wilson, R. W., et al. 1987, \apjs~65, 13

\bibitem[Bally et al. 1988]{}
Bally, J., Stark, A. A., Wilson, R. W., et al. 1988, \apj~324, 223

\bibitem[Benjamin et al. 2005]{}
Benjamin, R. A., Churchwell, E., Babler, B. L., et al. 2005, \apjl~630, L149

\bibitem[Binney et al. 1991]{}
Binney, J., Gerhard, O. E., Stark, A. A., et al. 1991, \mnras~252, 210

\bibitem[Binney et al. 1997]{}
Binney, J., Gerhard, O., \& Spergel, D. 1997, \mnras~288, 365

\bibitem[Binney and Merrifield 1998]{}
Binney, J., \& Merrifield, M. 1998,
in Galactic Astronomy,
ed. J. P. Ostriker and D. N. Spergel (Princeton, New Jersey)

\bibitem[Bitran 1987]{}
Bitran, M. 1987, Ph.D. thesis, University of Florida

\bibitem[Bitran et al. 1997]{}
Bitran, M., Alvarez, H., Bronfman, L., et al. 1997, \aas~125, 99

\bibitem[Bissantz et al. 2003]{}
Bissantz, N., Englmaier, P., \& Gerhard, O. 2003, \mnras~340, 949

\bibitem[Bissantz and Gerhard 2002]{} 
Bissantz, N., \& Gerhard, O. 2002, \mnras~330, 591
	
\bibitem[Burton and Liszt 1978]{}
Burton, W. B., \& Liszt, H. S. 1978, \apj~225, 815

\bibitem[Burton and Liszt 1992]{}
Burton, W. B., \& Liszt, H. S. 1992, \aas~95, 9

\bibitem[Burton and Liszt 1993]{}
Burton, W. B., \& Liszt, H. S. 1993, \aa~274, 765

\bibitem[Carr et al. 2000]{}
Carr, J. S., Sellgren, K., \& Balachandran, S. C. 2000, \apj~530, 307

\bibitem[Combes 1991]{}
Combes, F. 1991, \araa~29, 195

\bibitem[Contopoulos and Mertzanides 1977]{}
Contopoulos, G., \& Mertzanides, C. 1977, \aa~61, 477

\bibitem[Contopoulos and Papayannopoulos 1980]{}
Contopoulos, G., \& Papayannopoulos, Th. 1980, \aa~92, 33

\bibitem[Cordes and Lazio 2002]{}
Cordes, J. M., \& Lazio, T. J. W. 2002, {\it astro-ph/0207156}

\bibitem[Cordes  et al. 1991]{}
Cordes, J. M., Weisberg, J. M., Frail, D. A., et al. 1991, Nature~354, 121
	
\bibitem[Dahmen et al. 1998]{}
Dahmen, G., H\"uttemeister, S., Wilson, T. L., et al. 1998, \aa~331, 959

\bibitem[Cox 2000]{}
D\"appen, W. 2000,
in Allen's Astrophysical Quantities,
ed. A. N. Cox (Springer-Verlag, New York), 27

\bibitem[Ferriere 1998]{}
Ferri\`ere, K. M. 1998, \apj~497, 759

\bibitem[Ferriere 2001]{}
Ferri\`ere, K. M. 2001, \rvmp~73, 1031

\bibitem[Fux 1999]{}
Fux, R. 1999, \aa~345, 787

\bibitem[Englmaier and Gerhard 1999]{}
Englmaier, P., \& Gerhard, O. 1999, \mnras~304, 512

\bibitem[Heiligman 1987]{}
Heiligman, G. M. 1987, \apj~314, 747

\bibitem[Jenkins and Binney 1994]{}
Jenkins, A., \& Binney, J. 1994, \mnras~270, 703

\bibitem[Kaifu et al. 1972]{}
Kaifu, N., Kato, T., \& Iguchi, T. 1972, Nature 238, 105

\bibitem[Koyama et al. 1989]{}
Koyama, K., Awaki, H., Kunieda, H., et al. 1989, Nature~339, 603
	
\bibitem[Koyama et al. 1996]{}
Koyama, K., Maeda, Y., Sonobe, T., et al. 1996, Publ. Astron. Soc. Japan~48, 249

\bibitem[Launhardt et al. 2002]{}
Launhardt, R., Zylka, R., \& Mezger, P. G. 2002, \aa~384, 112

\bibitem[Lazio and Cordes 1998]{}
Lazio, T. J. W. \& Cordes, J. M. 1998, \apj~505, 715

\bibitem[Liszt and Burton 1978]{}
Liszt, H. S., \& Burton, W. B. 1978, \apj~226, 790

\bibitem[Liszt and Burton 1980]{}
Liszt, H. S., \& Burton, W. B. 1980, \apj~236, 779

\bibitem[Liszt and Burton 1996]{}
Liszt, H. S., \& Burton, W. B. 1996, 
in Unsolved problems of the Milky Way, 
Proceedings of the 169th IAU Symposium, 
ed. L. Blitz and P. Teuben (Kluwer, Dordrecht), 297

\bibitem[Maciel and Quireza 1999]{}
Maciel, W. J., \& Quireza, C. 1999, \aa~345, 629

\bibitem[Magnani et al. 2006]{}
Magnani, L., Zelenik, S., Dame, T. M., et al. 2006, \apj~636, 267

\bibitem[Martin et al. 2004]{}
Martin, C. L., Walsh, W. M., Xiao, K., et al. 2004, \apjs~150, 239

\bibitem[Mehringer et al. 1992]{}
Mehringer, D. M., Yusef-Zadeh, F., Palmer, P., et al. 1992, \apj~401, 168

\bibitem[Mehringer et al. 1993]{}
Mehringer, D. M., Palmer, P., Goss, W. M., et al. 1993, \apj~412, 684

\bibitem[Mezger et al. 1996]{}
Mezger, P. G., Duschl, W. J., \& Zylka, R. 1996, \aar~7, 289
	
\bibitem[Mezger and Pauls 1979]{}
Mezger, P. G., \& Pauls, T. 1979, 
in The Large-Scale Characteristics of the Galaxy,
Proceedings of the 84th IAU Symposium,
ed. W. B. Burton (Reidel, Dordrecht), 357

\bibitem[Morris and Serabyn 1996]{}
Morris, M., \& Serabyn, E. 1996, \araa~34, 645

\bibitem[Najarro et al. 2004]{}
Najarro, F., Figer, D. F., Hillier, D. J., et al. 2004, \apjl~611, L105

\bibitem[Oka et al. 1998]{}
Oka, T., Hasegawa, T., Sato, F., et al. 1998, \apj~493, 730
	
\bibitem[Oka et al. 2005]{}
Oka, T., Geballe, T. R., Goto, M., et al. 2005, \apj~632, 882
	
\bibitem[Rodr\'iguez-Fern\'andez et al. 2001]{}
Rodr\'iguez-Fern\'andez, N. J., Mart\'in-Pintado, J., Fuente, A., 
et al. 2001, \aa~365, 174

\bibitem[Rohlfs and Braunsfurth 1982]{}
Rohlfs, K., \& Braunsfurth, E. 1982, \aa~113, 237

\bibitem[Rolleston et al. 2000]{}
Rolleston, W. R. J., Smartt, S. J., Dufton, P. L., et al. 2000, \aa~363, 537

\bibitem[Sanders et al. 1984]{}
Sanders, D. B., Solomon, P. M., \& Scoville, N. Z. 1984, \apj~276, 182

\bibitem[Sawada et al. 2004]{}
Sawada, T., Hasegawa, T., Handa, T., et al. 2004, \mnras~349, 1167

\bibitem[Scoville 1972]{}
Scoville, N. Z. 1972, \apjl~175, L127

\bibitem[Sellwood and Sparke 1988]{}
Sellwood, J. A., \& Sparke, L. S. 1988, \mnras~231, 25

\bibitem[Shaver et al. 1983]{}
Shaver, P. A., McGee, R. X., Newton, L. M., et al. 1983, \mnras~204, 53

\bibitem[Smartt et al. 2001]{}
Smartt, S. J., Venn, K. A., Dufton, P. L., et al. 2001, \aa~367, 86

\bibitem[Snowden et al. 1997]{}
Snowden, S. L., Egger, R., Freyberg, M. J., et al. 1997, \apj~485, 125

\bibitem[Sodroski et al. 1995]{}
Sodroski, T. J., Odegard, N., Dwek, E., et al. 1995, \apj~452, 262

\bibitem[Sofue 1995a]{}
Sofue, Y. 1995a, \pasj~47, 527

\bibitem[Sofue 1995b]{}
Sofue, Y. 1995b, \pasj~47, 551

\bibitem[Stark et al. 1988]{}
Stark, A. A., Bally, J., Knapp, G. R., et al. 1988, 
in Molecular Clouds in the Milky Way and External Galaxies,
ed. R. L. Dickman, R. L. Snell, and J. S. Young (Springer-Verlag, Berlin), 303

\bibitem[Strong et al. 2004]{}
Strong, A. W., Moskalenko, I. V., Reimer, O. et al. 2004, \aa~422, L47

\bibitem[Taylor and Cordes 1993]{}
Taylor, J. H., \& Cordes, J. M. 1993, \apj~411, 674

\bibitem[Wada 1994]{}
Wada, K. 1994, PASJ~46, 165

\bibitem[Wolfire et al. 1995]{}
Wolfire, M. G., McKee, C. F., Hollenbach, D., et al. 1995, \apj~453, 673

\bibitem[Yamauchi et al. 1990]{}
Yamauchi, S., Kawada, M., Koyama, K., et al. 1990, \apj~365, 532

\end{thebibliography}
\end{document}